%% file: main.tex
\documentclass[12pt,onecolumn,amsmath,amssymb, spacing=1.5pt]{article}
\usepackage[letterpaper,margin=1in]{geometry}
\usepackage{ulem}

\newcommand{\bd}{\boldsymbol{d}}
\newcommand{\bc}{\boldsymbol{c}}
\newcommand{\bff}{\textbf{f}}

\newcommand{\bq}{\boldsymbol{q}}
\newcommand{\br}{\boldsymbol{\textbf{r}}}

\newcommand{\bt}{\boldsymbol{t}}

\newcommand{\bx}{\boldsymbol{\textbf{x}}}

\newcommand{\bR}{\boldsymbol{\textbf{R}}}
\newcommand{\bS}{\boldsymbol{\textbf{S}}}
\newcommand{\bJ}{\boldsymbol{\textbf{J}}}

\newcommand{\by}{\boldsymbol{\textbf{y}}}

\newcommand{\dr}{\,d\br}

\newcommand{\vext}{v_{\rm{ext}}}

\newcommand{\Exc}{E_{\rm{xc}}}

\newcommand{\bsmear}{b_{\text{\footnotesize{s}}}}
\newcommand{\vsmear}{v_{\text{\footnotesize{s}}}}
\newcommand{\vaux}{v_{\text{\footnotesize{aux}}}}
\newcommand{\bunit}{b_{\text{\footnotesize{u}}}}
\newcommand{\EZZ}{E_{\text{\footnotesize{zz}}}}
\newcommand{\Eself}{E_{\text{\footnotesize{self}}}}

\newcommand{\EH}{E_{\text{\footnotesize{H}}}}
\newcommand{\Ftilde}{\tilde{F}}
\newcommand{\DTE}{\Delta {E}}
\newcommand{\TE}{\widetilde{E}}

\newcommand{\braket}[2]{\left< #1 \vphantom{#2} \right | \left. #2 \vphantom{#1} \right>} 

\newcommand{\norm}[1]{\left\lVert#1\right\rVert}

\usepackage{doi}
\usepackage{mathtools}
\usepackage{footnote}
\usepackage{hyperref}
\hypersetup{
    colorlinks=true,
    linkcolor=magenta,
    filecolor=magenta,      
    urlcolor=cyan,
    }

\usepackage{subfiles}
\usepackage{graphicx}
\usepackage{graphics}
\usepackage{multirow}
\usepackage{amsxtra}
\usepackage{stmaryrd}
\usepackage{mathrsfs}
\usepackage{caption}
\usepackage{subcaption}
\usepackage{epsfig}
\usepackage{rotating}
\usepackage{setspace}
\usepackage{float}
\usepackage{amsfonts}
\usepackage{amsbsy}
\usepackage{amscd}
\usepackage{amsthm}
\usepackage{relsize}
\usepackage{color}
\usepackage{tablefootnote}
\usepackage{tabularx}
\usepackage{caption}
\usepackage{sidecap}
\usepackage{dcolumn}
\usepackage{algorithm}
\usepackage{algorithmic}
\usepackage{physics}
\usepackage{authblk}

\definecolor{hellgruen}{rgb}{0.2,0.7,0.2}

\newcolumntype{M}[1]{>{\centering\arraybackslash}m{#1}}
\newcolumntype{N}{@{}m{0pt}@{}}
\begin{document}
\title{Field theoretic atomistics: Learning thermodynamic and variational surrogate to density functional theory}
\author[a, \dag, *]{Sambit Das}
\author[a, \dag, *] {Bikash Kanungo}
\author[a] {Arghadwip Paul}
\author[a,b]{Vikram Gavini}
\affil[a]{\small Department of Mechanical Engineering, University of Michigan, Ann Arbor, Michigan 48109, USA}
\affil[b]{\small Department of Materials Science and Engineering, University of Michigan, Ann Arbor, Michigan 48109, USA}
\affil[$\dag$]{These authors contributed equally}
\affil[$*$]{Corresponding authors: \{\texttt{dsambit, bikash}\}\texttt{@umich.edu}}
\date{}

\maketitle

\begin{abstract}
 The Hohenberg-Kohn (HK) theorem---the bedrock of density functional theory (DFT)---establishes a universal map from the external potential to the  energy. It also relates the electron density and atomic forces to the variation of the energy with the external potential. 
 But the HK map is rarely utilized in atomistics, wherein interatomic potentials are defined using the molecular or crystal structure rather than the external potential. As a break from this tradition, we present a \textit{field theoretic} atomistics framework where the external potential assumes the central quantity. We machine learn the HK energy map while satisfying the thermodynamic limit. Further, we obtain both forces and electron density from the variation of the HK energy map, that are exact relations. Our models attain good accuracy across diverse benchmarks and compete with state-of-the-art machine learned interatomic potentials. Through electron density, we predict accurate dipole and quadrupole moments, otherwise nontrivial for interatomic potentials. Our formulation paves the way for a scalable electronic structure surrogate to DFT.                
\end{abstract}

\input{intro}
\input{results}
\input{discussion}
\input{method}

\section*{Acknowledgments}
We acknowledge the use of the Great Lakes HPC Cluster at University of Michigan. Additionally, this research also used resources of the National Energy Research Scientific Computing Center, a DOE Office of Science User Facility supported by the Office of Science of the U.S. Department of Energy under Contract No. DE-AC02-05CH11231. 

\paragraph{Author contributions:} S.D. and B.K. conceptualized the idea and led the research, with crucial inputs from V.G. A.P. helped with optimizing the FTA code. All authors contributed towards writing the manuscript.

\paragraph{Competing interests:} Authors declare that they have no competing interests.

\paragraph{Data availability:} The authors declare that all the data supporting the results of this study are available upon reasonable request to the corresponding author.

\paragraph{Code availability:} The code used to machine learn FTA models is available upon reasonable request to the corresponding author.

\clearpage
\bibliography{ref}
\bibliographystyle{unsrt}

\pagebreak
\begin{center}
\textbf{\Large Supplementary Information}
\end{center}

\setcounter{equation}{0}
\setcounter{figure}{0}
\setcounter{table}{0}
\setcounter{page}{1}
\setcounter{section}{0}
\makeatletter
\renewcommand{\theequation}{S\arabic{equation}}
\renewcommand{\thefigure}{S\arabic{figure}}
\renewcommand{\thetable}{S\arabic{table}}
\renewcommand{\thesection}{S\arabic{section}}
\renewcommand{\thepage}{S-\arabic{page}}

\input{SI}

\end{document}

%% file: intro.tex
\section{Introduction} \label{sec:label}
Density functional theory (DFT)~\cite{Hohenberg1964, Kohn1965,Jones2015, Burke2012} and interatomic potentials (IP)~\cite{Frenkel2023understanding, Muser2023interatomic, Farah2012classical, Behler2016perspective, Mishin2021machine, Deringer2019machine, Kocer2022neural}  have long remained indispensable tools in computational chemistry and materials science. Although both are intricately tied to the principles of quantum mechanics, DFT and IPs have evolved as two disparate methods. DFT is applied to understand the electron-electron and electron-nuclear interactions and IPs are used to define the potential energy surface (PES) for the atoms in terms of the molecular or crystal structure. Thus, while DFT is an inherently electronic structure method, the IPs are considered to be atomistic models without any knowledge or predictability of the electronic structure. This distinction, however, is artificial, as theoretically both DFT and IPs are conjoined by the Hohenberg-Kohn (HK) theorem~\cite{Hohenberg1964}. The HK theorem establishes the groundstate electronic energy ($E_e$) as a universal functional of the external potential, denoted as $\vext(\br)$. That is, as per the HK theorem, $E_e \equiv E_e[\vext]$. The external potential can represent the nuclear potential and/or any externally applied potential. The PES ($E$) can be obtained by adding the nuclear-nuclear repulsion energy ($\EZZ$) to $E_e$. The HK theorem also proves a one-to-one map between the groundstate electron density ($\rho_g(\br)$) and the external potential, given as $\rho_g(\br) = \frac{\delta E_e[\vext]}{\delta \vext(\br)}$. As a result, defining the PES in terms of $\vext$, as opposed to the molecular or crystal structure, permits the evaluation of the density, through the variation of $E_e$ with $\vext$. In other words, the HK theorem allows us to model IPs that, in principle, can be at the same footing as DFT and provide insights into the electronic structure of matter.

\noindent This prowess of the HK theorem has seldom been utilized while designing IPs. Two recent efforts~\cite{Brockherde2017bypassing, Shao2023machine} have pioneered the idea of using the HK theorem to, in spirit, design IPs. However, these models have crucial limitations that can affect their practical utility. First, they do not maintain thermodynamic consistency: they do not satisfy the thermodynamic limit of the energy per atom being a bounded constant for an extended system (i.e., when number of atoms tend to infinity). Second, they do not satisfy the variational consistency of the density and forces being related to the variation of $E_e$ with respect to $\vext(\br)$: $\rho_g(\br) = \frac{\delta E_e[\vext]}{\delta \vext(\br)}$ and $\bff_I = \int \rho_g(\br) \nabla_{\bR_I}\vext(\br)\dr$. 
This is a subtle aspect linked to the fact that the previous efforts model $E_e[\vext]$ as a two-step map, first from $\vext(\br)$ to $\rho_g(\br)$ (or $\gamma(\br,\br')$, the one-electron reduced density matrix (1RDM)) and then from $\rho$ (or $\gamma$) to $E_e$, which breaks the variational consistency.      

\noindent In this work, we present an HK map that satisfies the thermodynamic limit and is variationally consistent in the density and atomic forces. This entails a reformulation of the HK map in terms of auxiliary quantities, and subsequently, machine learning of the HK map in terms of features of the auxiliary quantities. We term this approach as \textit{field theoretic atomistics} (FTA), as it inherently deals with fields as its inputs and outputs (e.g,, $\vext(\br)$ and $\rho_g(\br)$). This is in contrast to IPs which deal with structure as input and scalars (energy, forces) as outputs. The FTA approach offers several key advantages over IPs. First, it handles both nuclear potential and any externally applied potential (e.g., from an electric field) on equal footing. Second, it does not require any chemical species embedding, and thereby, avoids any combinatorial complexities that arise in IPs while dealing with multiple chemical species. We demonstrate that for widely used benchmarks our FTA models compete with state-of-the-art machine-learning IPs (MLIPs) in energy and force, while simultaneously predicting accurate densities and their dipole and quadrupole moments. This highlights the promise of the FTA models to serve as an accurate and scalable surrogate to DFT.        
\begin{figure}[hbt!]
    \centering
    \includegraphics[scale=0.30]{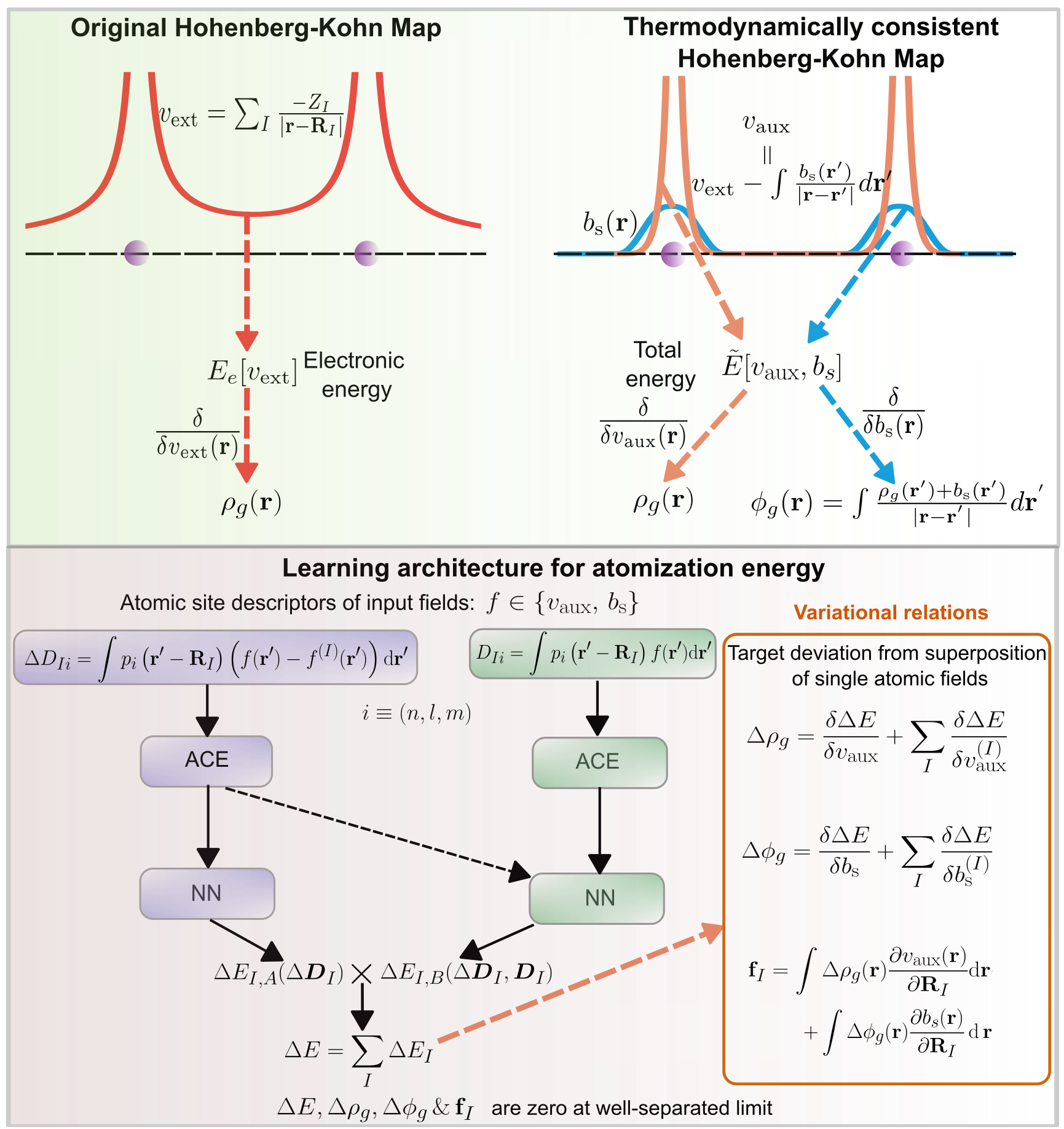}
    \caption{Schematic of field theoretic atomistics (FTA) approach. The original Hohenberg-Kohn (HK) map pertains to the electronic energy, which does not satisfy the thermodynamic limit. We reformulate the HK map for the total energy, in terms of auxiliary fields ($\vaux(\br)$, $\bsmear(\br)$), to ensure the thermodynamic limit. The variation of the energy with the auxiliary fields provide the ground-state density and electrostatic potential. We machine-learn the FTA energy map using a combination of atomic cluster expansion~\cite{Drautz2019} and neural networks.}
 \label{fig:schematicFTA}
\end{figure}


%% file: results.tex
\section{Results} \label{sec:results}
\subsection*{Hohenberg-Kohn theorem to Field Theoretic Atomistics}
We begin with the variational problem in DFT, where given an external potential ($\vext(\br)$) and number of electrons ($N_e$), the groundstate electronic energy $E_e$ can be obtained as
\begin{equation}\label{eq:DFTVar}
    E_e[\vext, N_e] = \min_{\rho \in A_N} \left(F[\rho] + \int \rho(\br) \vext(\br)\dr \right)\,, \quad \text{~s.t.~} \int \rho(\br) = N_e\,,
\end{equation}
where $A_N$ denotes the space of $N$-representable densities (i.e., densities corresponding to an antisymmetric wavefunction). $F[\rho]$ is a universal density functional, comprising of three components, $F[\rho] = T_s[\rho] + \Exc[\rho] + \EH[\rho]$, where $T_s[\rho]$, $\Exc[\rho]$, and $\EH[\rho]$ are the non-interacting kinetic energy,  exchange-correlation energy, and Hartree energy, respectively. While $T_s$ and $\Exc$ are known to be universal functionals of $\rho$, their forms are unknown, thereby requiring approximations. The Hartree energy, which represents the classical electrostatics interaction between electrons, is given as $\EH[\rho] = \frac{1}{2}\int \int \frac{\rho(\br)\rho(\br')}{|\br-\br'|}\dr\dr'$.  Typically, $\vext(\br)$ is the nuclear potential: $\vext(\br) = - \sum_{I}^{N_a} \frac{Z_I}{|\bR_I-\br|}$, where $N_a$ is the number of atoms and $Z_I$ and $\bR_I$ are the atomic number and position of the $I$-th atom. The PES (or total groundstate energy) includes the nuclear-nuclear repulsion: $E = E_e[\vext,N_e] + \EZZ$, where $\EZZ = \frac{1}{2}\sum_{I}\sum_{J\neq I}\frac{Z_IZ_J}{|\bR_I-\bR_J|}$. We note that while $E$ satisfies the thermodynamic limit, $E_e$ and $\EZZ$ individually do not. $E_e$ contains $\EH[\rho]$ and $\int \rho\vext\dr$, where both being electrostatic interactions grow quadratically with number of atoms, and hence, $E_e$ does not satisfy the thermodynamic limit. Similar arguments apply to the violation of thermodynamic limit for $\EZZ$. Thus, for thermodynamic consistency, we need to reformulate the variational problem in Eq.~\eqref{eq:DFTVar} in terms of $E$. To do so, we introduce a superposition of non-overlapping auxiliary charges on each nucleus: $\bsmear(\br) = \sum_I b_I(\br)$, where $ b_I(\br) = -Z_I \bunit(|\br-\bR_I|; r_c)$. Here $\bunit(r, r_c)$ is a compactly supported positive unit charge distribution with a cutoff radius of $r_c$ (refer to Sec. S3 in the SI for the form of $\bunit$). 
The $r_c$ satisfies $2r_c < \min_{IJ}|\bR_I-\bR_J|$ to ensure $b_I$'s do not overlap. It is straightforward to show that $\EZZ = \frac{1}{2}\sum_I\sum_{J\neq I}\int \int \frac{b_I(\br)b_J(\br')}{|\br-\br'|}\dr\dr'$. We also define an auxiliary potential $\vaux(\br) = \vext(\br) - \vsmear(\br)$, where $\vsmear(\br) = \int \frac{\bsmear(\br')}{|\br-\br'|}\dr'$.
We can define the PES as
\begin{equation} \label{eq:DFTVar2}
\begin{split}
    E[\vaux, \bsmear, N_e] &= \min_{\rho \in A_N} \left(\Ftilde[\rho] + \EH[\rho] + \int \rho(\br) \vaux(\br)\dr + \int \rho(\br) \vsmear(\br) \dr\right)\\
    &+ \EZZ\,, \quad \text{s.t.}~ \int \rho(\br)\dr = N_e\,,
\end{split}
\end{equation}
where $\Ftilde[\rho] = T_s[\rho] + \Exc[\rho]$. We note that $\EH[\rho] + \int \rho(\br)\vsmear(\br)\dr + \EZZ$ represents electrostatic energy of $\rho + \bsmear$, which can be reformulated as the following variational problem~\cite{Gavini2007}:  
\begin{equation} \label{eq:es}
 \EH[\rho] + \int \rho(\br)\vsmear(\br)\dr + \EZZ= \max_{\phi} \underbrace{\left(-\frac{1}{8\pi}\int |\nabla\phi(\br)|^2\dr + \int \left(\rho(\br)+\bsmear(\br)\right)\phi(\br)\dr\right)}_{P[\phi, \rho, \bsmear]} - \Eself\,,
\end{equation}
The maximizer of the functional $P[\phi, \rho, \bsmear]$, say $\phi^*(\br)$, corresponds to the electrostatic potential of $\rho + \bsmear$, given by $\phi^*(\br) = \int \frac{\rho(\br')+ \bsmear(\br')}{|\br-\br'|}\dr'$. Further, the maxima of $P[\phi, \rho, \bsmear]$  corresponds to the electrostatic energy of $\rho + \bsmear$ along with the self-interaction of $b_I$'s. That is, $P[\phi^*, \rho, \bsmear] = \frac{1}{2}\int \int \frac{\left(\rho(\br)+\bsmear(\br)\right)\left(\rho(\br')+\bsmear(\br')\right)}{|\br-\br'|}\dr\dr'$. To recover the correct electrostatic energy, we subtract out from $P[\phi^*, \rho, \bsmear]$ the spurious self-interaction of $b_I$'s, given by $\Eself=\sum_I \int \int \frac{b_I(\br) b_I(\br')}{|\br-\br'|}\dr\dr'$. 
Now, combining Eq.~\eqref{eq:DFTVar2} and Eq.~\eqref{eq:es}, we get the following saddle point (min-max) problem,
\begin{equation} \label{eq:DFTSaddle}
    \TE[\vaux, \bsmear, N_e] = \min_{\rho\in A_N} \max_{\phi} \left(\Ftilde[\rho] + \int \rho(\br) \vaux(\br)\dr + P[\phi, \rho, \bsmear]\right), \, \text{s.t.}\, \int \rho(\br)\dr = N_e\,,
\end{equation}
where $\TE[\vaux, \bsmear, N_e] = E[\vaux, \bsmear, N_e] + \Eself$. We note that $\Eself$ is independent of the minimization over $\rho$ and $\phi$, and is determined only by the form of the unit charge $\bunit(r,r_c)$. Thus, given a model for $\TE[\vaux, \bsmear, N_e]$, the PES ($E$) can be trivially obtained by subtracting out the $\Eself$. One can recast Eq.~\eqref{eq:DFTSaddle} as an unconstrained optimization by introducing a Lagrange multiplier ($\mu$) for the constraint $\int \rho(\br)\dr=N_e$. Solving the saddle point problem gives the groundstate solution, which we denote as $\rho_g$, $\phi_g$, and $\mu_g$. So far, we have replaced the canonical variational problem in DFT over $E_e$ (which does not satisfy the thermodynamic limit) with a saddle point problem over $\title{E}$ (which satisfies the thermodynamic limit). This saddle point reformulation of DFT, or variations of this, has been employed in various real-space DFT methods~\cite{Pask1999, Pask2005, Suryanarayana2010, Motamarri2012, Motamarri2013, Rufus2021, MOTAMARRI2020106853, das2022dft} as a means to have a unified framework for both finite (molecules, clusters) and extended systems (solids). 

\noindent Here, we show that the above saddle point reformulation results in various bijective maps and variational relations similar to the HK theorem. These bijective maps and variational relations are (refer to Sec. S1 in the SI for the proof):
\begin{equation}\label{eq:bijective}
\left(\rho_g(\br), \mu_g\right) \leftrightarrow \left(\vaux(\br), N_e\right);~\text{and} \qquad \phi_g(\br) \leftrightarrow \bsmear(\br)
\end{equation}
\begin{equation}\label{eq:slope}
    \frac{\delta \TE[\vaux, \bsmear, N_e]}{\delta \vaux(\br)} = \rho_g(\br);\qquad \frac{\delta \TE[\vaux, \bsmear, N_e]}{\delta \bsmear(\br)} = \phi_g(\br); ~\text{and} \qquad \frac{\partial \TE[\vaux, \bsmear, N_e]}{\partial N_e} = \mu_g\,.
\end{equation}
Using the above variational relations,  atomic force $\bff_I$ evaluates to 
\begin{equation} \label{eq:force}
    \bff_I = \int \rho_g(\br) \frac{\partial \vaux(\br)}{\partial \bR_I}\dr + \int \phi_g \frac{\partial \bsmear(\br)}{\partial \bR_I}\dr\,. 
\end{equation}
The above are the key results that lay the foundation of our FTA models. We remark that, the above FTA formulation does not assume any specific form for the universal density functional $F[\rho]$, and hence, is valid  when used with any approximate XC functional in DFT. More importantly, the FTA formulation can be applied to any other electronic structure method (e.g., configuration interaction (CI), coupled-cluster (CC), quantum Monte Carlo (QMC), etc.), as different electronic structure method would merely alter the form of $F[\rho]$. Equations~\eqref{eq:DFTSaddle}-\eqref{eq:slope} allow us to machine learn $\TE$, thereby allowing for thermodynamically and variationally consistent density and forces. Beyond serving as exact theoretical conditions, the variational consistency leads to important symmetry relations.  A key physically relevant relation is that $\rho_g$ and $\phi_g$  transform covariantly under  energy invariant transformations of the input fields ($\vaux$ and $\bsmear$) in the $\TE$ functional.  To elaborate, let $T$ denote the representation of the Euclidean group (rotations, translations, inversions, reflections)~\cite{geiger2022e3nn} that transforms $\vaux$ as $\vaux^{\prime}= T \vaux$ and $\bsmear$ as $\bsmear^{\prime}= T \bsmear$. Using the energy invariance we can write
    $\TE[\vaux^{\prime}, \bsmear^{\prime}, N_e]=\TE[\vaux, \bsmear, N_e]$,
and subsequently analyzing the behavior of the differential on both sides with respect to $\vaux$ we obtain the covariant transformation of $\rho_g$:
\begin{align}\label{eq:covariantTransformationDensity}
T^{\dagger} \rho_g[T\vaux,T\bsmear] = \rho_g[\vaux,\bsmear]\,, \quad T^{\dagger} \phi_g[T\vaux,T\bsmear] = \phi_g[\vaux,\bsmear]\,,
\end{align}
where $T^{\dagger}$ denotes the adjoint of $T$. We refer to Sec. 1.3 in the SI for the derivation of the above relation. 
We note that when $T$ is an unitary operator ($T={\left(T^{\dagger}\right)}^{-1}$), such as rotation, the transformation in Eq.~\ref{eq:covariantTransformationDensity} simplifies to an equivariance
\begin{equation}
    \rho_g[T\vaux,T\bsmear]= T \rho_g[\vaux,\bsmear]\,, \quad     \phi_g[T\vaux,T\bsmear]= T \phi_g[\vaux,\bsmear]\,.
\end{equation}
In other words, applying the group operation to the inputs ($\vaux,\bsmear$) transforms the output ($\rho_g, \phi_g$) in a similar way.   
Another symmetry relation that manifests from  $\rho_g$ and $\phi_g$ being functional derivatives of an energy functional is the mixed derivative condition: 
\begin{align}\label{eq:rhogVariationalCorrolarryProperties}
  \frac{ \delta \rho_g(\br)}{\delta \vaux(\br^{\prime})}=\frac{ \delta \rho_g(\br^{\prime})}{\delta \vaux(\br)}\,,\qquad 
  \frac{ \delta \phi_g(\br)}{\delta \bsmear(\br^{\prime})}=\frac{ \delta \phi_g(\br^{\prime})}{\delta \bsmear(\br)}\,,
\end{align} 
which can be easily verified by substitution of the variational relations for  $\rho_g$ and $\phi_g$ (cf. Eq.~\ref{eq:slope}) in the above expressions.  We remark that machine-learning frameworks have been  developed  for direct prediction of $\rho_g$ from atomic-structure  that satisfy the covariant transformation condition~\cite{fu2024recipe,koker2024higher}, but they do not satisfy the mixed derivative condition. The advantage of $\rho_g$  prediction in the FTA approach is that the mixed derivative condition is satisfied independent of the learning architecture, and further the covariant transformation condition requires only the  energy to be invariant with respect to the Euclidean group transformation. 


\noindent Modeling $\TE$ that can accurately predict $\rho_g$ and $\phi_g$ (via Eq.~\ref{eq:slope}) requires learning the sharp features in the density and the electrostatic potential, especially near the nuclei. In contrast, the difference of $\rho_g$ ($\phi_g$) from the superposition of atomic densities (atomic electrostatic potentials) is a much smoother field, and hence, more amenable to learning. Further, much of chemical bonding is mediated by these differences in densities and electrostatic potentials. We, therefore, extend the above analysis to the atomization energy 
\begin{equation} \label{eq:AE}
\DTE[\vaux, \bsmear, N_e, \{\vaux^{(I)}\}, \{\bsmear^{(I)}\}, \{N_e^{(I)}\}]  = \TE[\vaux, \bsmear, N_e] - \sum_I \TE[\vaux^{(I)}, \bsmear^{(I)}, N_e^{(I)}]\,,  
\end{equation}
where $\vaux^{(I)}$ and $\bsmear^{(I)}$ are the $\vaux$ and $\bsmear$ for the $I$-th isolated atom; and $N_e^{(I)}=Z_I$ (neutral atom). We emphasize that, in the above, although $\vaux$ has an apparent dependence on $\vaux^{(I)}$, both are independent fields, as they are inputs to different energy functionals: $\vaux$ to the molecular energy and $\vaux^{(I)}$ to the isolated atom energy for the $I$-th atom. Similarly, $\bsmear$ and $\bsmear^{(I)}$ are independent fields. Using Eq.~\ref{eq:slope}, it is straightforward to note that 
\begin{equation} \label{eq:deltaSlope}
    \frac{\delta \DTE}{\delta \vaux} + \sum_I \frac{\delta \DTE}{\delta \vaux^{(I)}} = \Delta \rho_g; \quad     \frac{\delta \DTE}{\delta \bsmear} + \sum_I \frac{\delta \DTE}{\delta \bsmear^{(I)}} = \Delta \phi_g;~ \textrm{and}  \quad     \frac{\partial \DTE}{\partial N_e} =  \mu_g\,.
\end{equation}
In the above, $\Delta \rho_g$ is the difference between $\rho_g$ and the superposition of atomic densities: $\Delta \rho_g = \rho_g - \sum_I \rho_g^{(I)}$, with $\rho_g^{(I)}$ being the groundstate density of the $I$-th isolated atom. $\Delta \phi_g$ is defined similarly. As is customary in atomistic models, we invoke the atomic partitioning \textit{ansatz} to write $\DTE = \sum_I\DTE_I$, where $\DTE_I$ denotes the contribution from $I$-th atom. Theoretically, $\DTE_I$ depends on all the inputs to $\DTE$, namely, the total quantities ($\vaux$, $\bsmear$, $N_e$) and the set of isolated atom quantities ($\{\vaux^{(I)}\}$, $\{\bsmear^{(I)}\}$, and $\{N_e^{(I)}\}$). However, as a simplification, we approximate $\DTE_I$ to depend only the total quantities ($\vaux$, $\bsmear$, $N_e$) and the self-quantities ($\vaux^{(I)}$, $\bsmear^{(I)}$, $N_e^{(I)}$), i.e., we ignore the self-quantities from other atoms. This is not merely a matter of convenience, but motivated by the fact that the relevant chemistry around a nuclei can be modeled through the change in the local environment, which can be encapsulated through the total- and the self-quantities. So far, the FTA formulation along with the resulting atomization energy model of Eq.~\ref{eq:AE} has been presented in the context of local external potential, which typically pertains to all-electron DFT. However, the bulk of DFT calculations use the pseudopotential approximation, especially nonlocal pseudopotentials, for numerical efficiency. The above atomization energy model with its variational relations to $\Delta \rho_g$ and $\Delta \phi_g$ permits the use of reference data from pseudopotential DFT calculations. This is because both $\Delta \rho_g$ and $\Delta \phi_g$ from  pseudopotential calculations are expected to be close to their counterparts from all-electron DFT. Thus, one can use the above atomization energy formulation with all-electron $\vaux$ and $\bsmear$ but with reference $\Delta \rho_g$ and $\Delta \phi_g$ from pseudopotential calculations. In this work, we use all-electron DFT data as our reference.

\subsection*{Learning Field Theoretic Atomistics models}
We model $\DTE_I$ in terms of  the atom-centered descriptors of $\vaux, \bsmear, \vaux^{(I)}, \bsmear^{(I)}$. As a first attempt at FTA, we restrict our models to only charge-neutral systems. For charge-neutral systems the information of $N_e$ and $N_e^{(I)}$ is implicitly contained in $\bsmear$ and $\bsmear^{(I)}$, as $\int \bsmear(\br)\dr = N_e$ and $\int \bsmear^{(I)}(\br) \dr = N_e^{(I)}$. Thus, in our FTA models, we drop the explicit dependence on  $N_e$ and $N_e^{(I)}$. We model $\DTE_I[\vaux, \bsmear, N_e, \vaux^{(I)}, \bsmear^{(I)}, N_e^{(I)}]\approx \DTE_I[\bx_I, \by_I, \bx_I^{\text{self}}, \by_I^{\text{self}}]$, where  $\bx_I, \by_I, \bx_I^{\text{self}}, \by_I^{\text{self}}$ are the vectors containing the descriptors of $\vaux, \bsmear, \vaux^{(I)}, \bsmear^{(I)}$, given as
\begin{equation}\label{eq:descriptors}
\begin{split}
    &x_{i,I} = \int u_{i}(\br-\bR_I)\vaux(\br)\dr\,, \quad  x_{i,I}^{\text{self}} = \int u_{i}(\br-\bR_I)\vaux^{(I)}(\br)\dr\,,\\
    &y_{i,I} = \int w_{i}(\br-\bR_I)\bsmear(\br)\dr\,, \qquad  y_{i,I}^{\text{self}} = \int w_{i}(\br-\bR_I)\bsmear^{(I)}(\br)\dr\,.
\end{split}
\end{equation}
In the above, $u_i(\br)$'s are probe functions for $\vaux$ and $\vaux^{(I)}$; and $w_i(\br)$'s are the probe functions for $\bsmear$ and $\bsmear^{(I)}$. We consider $u_i$ and $w_i$ that take the form of atomic orbital:
\begin{equation}
u_i(\br)=f_n(r)\mathcal{Y}_{lm}(\theta,\phi)\,, \qquad w_i(\br)=g_n(r)\mathcal{Y}_{lm}(\theta,\phi)\,,   
\end{equation}
where $f_n$ and $g_n$ are compactly supported radial functions; $\mathcal{Y}_{lm}$ is the real form spherical harmonic with angular momentum $l$ and magnetic quantum number $m$; $r$, $\theta$, and $\phi$ denote the radial distance, polar angle and azimuthal angle of $\br$ from the origin, respectively; and the index $i\equiv(n,l,m)$. Writing $\DTE_I$ in terms of the above descriptors, we have 
\begin{equation}\label{eq:DEProd}
\begin{split}
    \DTE_I[\vaux, \bsmear, \vaux^{(I)}, \bsmear^{(I)}] &\approx \DTE_I[\bx_I, \by_I, \bx_I^{\text{self}}, \by_I^{\text{self}}]\,,
\end{split}
\end{equation}
where $\bx_I$ is the vector containing $x_{i,I}$ (similar definitions hold for $\bx_I^{\text{self}}$, $\by_I$, and $\by_I^{\text{self}}$).
Using Eq.~\ref{eq:deltaSlope}, the model $\Delta\rho_g$ and $\Delta \phi_g$ takes the  form
\begin{equation}\label{eq:deltaSlopeDeltaRho1}
\Delta \rho_g(\br) = \mathlarger{\mathlarger{\sum}}_I \left(\frac{\delta \DTE_I}{\delta \vaux(\br)} + \frac{\delta \DTE_I}{\delta \vaux^{(I)}(\br)}\right) = \mathlarger{\mathlarger{\sum}}_I \mathlarger{\mathlarger{\sum}}_i \left(\frac{\partial \DTE_I}{\partial x_{i,I}}+ \frac{\partial \DTE_I}{\partial x_{i,I}^{\text{self}}}\right)u_i(\br-\bR_I)\,, 
\end{equation}
\begin{equation}\label{eq:deltaSlopeDeltaRho2}
\Delta \phi_g(\br) = \mathlarger{\mathlarger{\sum}}_I \left(\frac{\delta \DTE_I}{\delta \bsmear(\br)} + \frac{\delta \DTE_I}{\delta \bsmear^{(I)}(\br)}\right) = \mathlarger{\mathlarger{\sum}}_I \mathlarger{\mathlarger{\sum}}_i \left(\frac{\partial \DTE_I}{\partial y_{i,I}}+ \frac{\partial \DTE_I}{\partial y_{i,I}^{\text{self}}}\right)w_i(\br-\bR_I)\,. 
\end{equation}
In other words, $u_i(\br-\bR_I)$ and $w_i(\br-\bR_I)$ serve as the basis to represent $\Delta \rho_g$ and $\Delta \phi_g$, respectively. Our choice for the radial probe functions ($f_n$ and $g_n$) is motivated by the above observation, i.e., they should serve as a basis for $\Delta \rho_g$ and $\Delta \phi_g$. In that light, we use a combination of short-ranged orthogonal Bessel functions and long-ranged damped-multipole functions for $f_n$ and $g_n$ (refer Sec. S2 in the SI for their forms). For well-separated atoms or an isolated atom, $\DTE$, $\Delta \rho_g$, and $\Delta \phi_g$ should all be zero. To satisfy this limit, we write $\DTE_I$ as
\begin{equation}\label{eq:separableForm}
    \DTE_I[\bx_I, \by_I, \bx_I^{\text{self}}, \by_I^{\text{self}}] = \DTE_{I,A}[\Delta\bx_I, \Delta \by_I]\,\, \DTE_{I,B}[ \Delta\bx_I, \Delta \by_I, \bx_I,  \by_I]\,, 
\end{equation}
where $\Delta\bx_I = \bx_I-\bx_I^{\text{self}}$; and $\Delta\by_I = \by_I-\by_I^{\text{self}}$. The above form allows us to satify the well-separated or isolated atoms limits by ensuring that $\DTE_{I,A}$  tends to zero when $\Delta \bx_I$ or $\Delta \by_I$ tend to zero. We model both $\DTE_{I,A}$ and $\DTE_{I,B}$ using a combination of atomic cluster expansion (ACE)~\cite{Drautz2019} and a feed-forward neural network (NN), the details of which are presented in the Methods section. We optimize the learnable parameters in our ACE and NN using the following loss function
\begin{equation} \label{eq:loss}
\begin{split}
    \mathcal{L} &= \frac{t_E}{M} \sum_{\alpha=1}^M \left(\DTE^{(\alpha)}_{\text{ref}}-\DTE^{(\alpha)}\right)^2 + \frac{t_F}{M}  \sum_{\alpha=1}^{M} \left[\sum_I \frac{1}{N_a}\left(\norm{\bff^{(\alpha)}_{I,\text{ref}}-\bff^{(\alpha)}_I}^2 \right) \right] \\ 
    & + \frac{t_{\rho}}{M} \sum_{\alpha=1}^{M}\norm{\Delta\rho_{g,\text{ref}}^{(\alpha)}(\br) - \Delta\rho_g^{(\alpha)}(\br)}_{\text{J}}^2
    + \frac{t_{\phi}}{M} \sum_{\alpha=1}^{M}\norm{\Delta\phi_{g,\text{ref}}^{(\alpha)}(\br) - \Delta\phi_g^{(\alpha)}(\br)}^2_{\text{S}}\,,
\end{split}
\end{equation}
where $\alpha$ indexes the $M$ training samples; $t_E$, $t_F$, $t_\rho$, and $t_\phi$ are the weights assigned to the energy, force, density, and potential loss terms; $\norm{\cdot}$ is the  Eucledian norm, $\norm{f(\br)}_{\text{S}} = \sqrt{\int (f(\br))^2\dr}$ is the $L_2$ norm for a continuous function; and $\norm{f(\br)}_{\text{J}} = \sqrt{\int \int \frac{f(\br)f(\br')}{|\br-\br'|}\dr\dr'}$ is the Coulomb norm. The use of the Coulomb norm is motivated by the fact that the Coulomb norm assigns a higher weight on the low angular momentum ($l$) of the density. As a result, it nudges the model towards better dipole and quadrupole moments of the density---quantities that are important to measure the response of a system to external electric field. For efficient evaluation of the field loss terms (density and potential loss in Eq.~\ref{eq:loss}), we represent the reference fields, $\Delta\rho_{g,\text{ref}}^{(\alpha)}(\br)$ and $\Delta\phi_{g,\text{ref}}^{(\alpha)}(\br)$ in the $u_i(\br)$ and $w_i(\br)$ probe function bases, respectively. Given that the predicted $\Delta \rho_g^{(\alpha)}(\br)$ and $\Delta \phi_g^{(\alpha)}(\br)$ can also be represented in the same probe function bases, one can greatly simplify the spatial integral in the field loss in terms of inexpensive matrix-vector products (see Sec. S5 in the SI for details).

\subsection*{rMD17-aspirin and 3BPA benchmarks}
In this section, we assess the accuracy of the proposed FTA approach using widely used benchmark molecular datasets: the aspirin molecule from the rMD17 dataset~\cite{rmd17}, and the 3BPA dataset~\cite{3BPA,3BPAMoleculeOriginal} (a flexible molecule with rotatable bonds). Both the benchmark systems have multiple chemical species, and hence, a hard test for the FTA approach that bypasses explicit chemical species embedding. We demonstrate that FTA successfully learns a combined model that does well on energy, forces, electron-density and electrostatic potential predictions.  We employ a learning architecture that combines ACE and NN on atom centered descriptors. Details of the datasets, learning architecture and training procedure are provided in the Methods section. We note that our focus here is a  proof-of-concept demonstration of the  key advantages and competence of the FTA approach. There is considerable scope for future research in applying FTA to more diverse benchmarks for molecular, surface and solid-state physics as well  development of better ML  architectures and training protocols that are theoretically well suited to FTA. 

\begin{figure}
    \centering
    \includegraphics[scale=0.7]{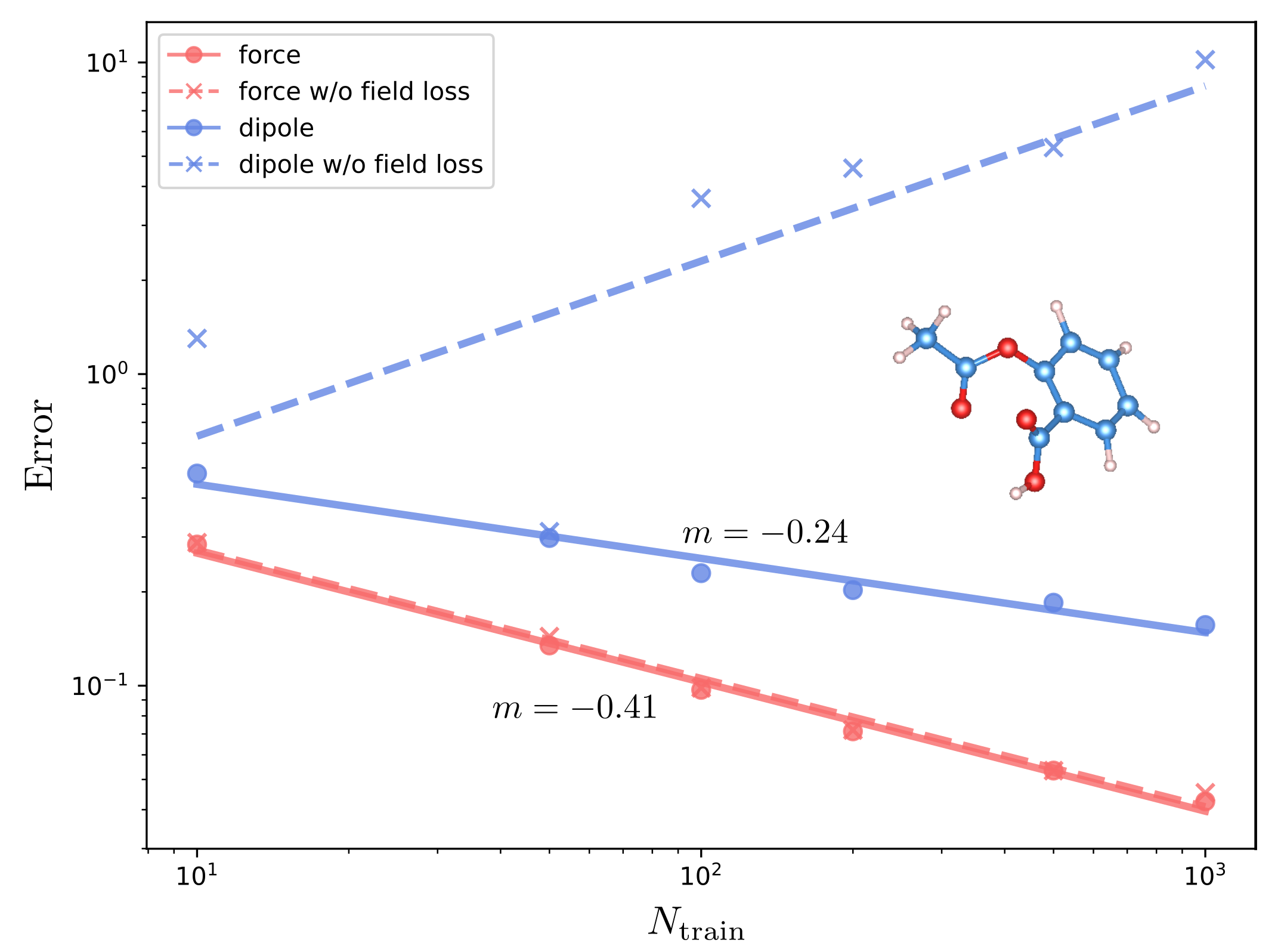}
    \caption{Learning curves of force and dipole moment for aspirin with training samples. For force, it shows the mean absolute error in meV/\AA~. For dipole moment, it shows the mean absolute error divided by the 
     mean of the absolute test set values (dimensionless quantity).}
    \label{fig:aspirin_learningcurve}
\end{figure}

\begin{figure}
    \centering
    \includegraphics[scale=0.6]{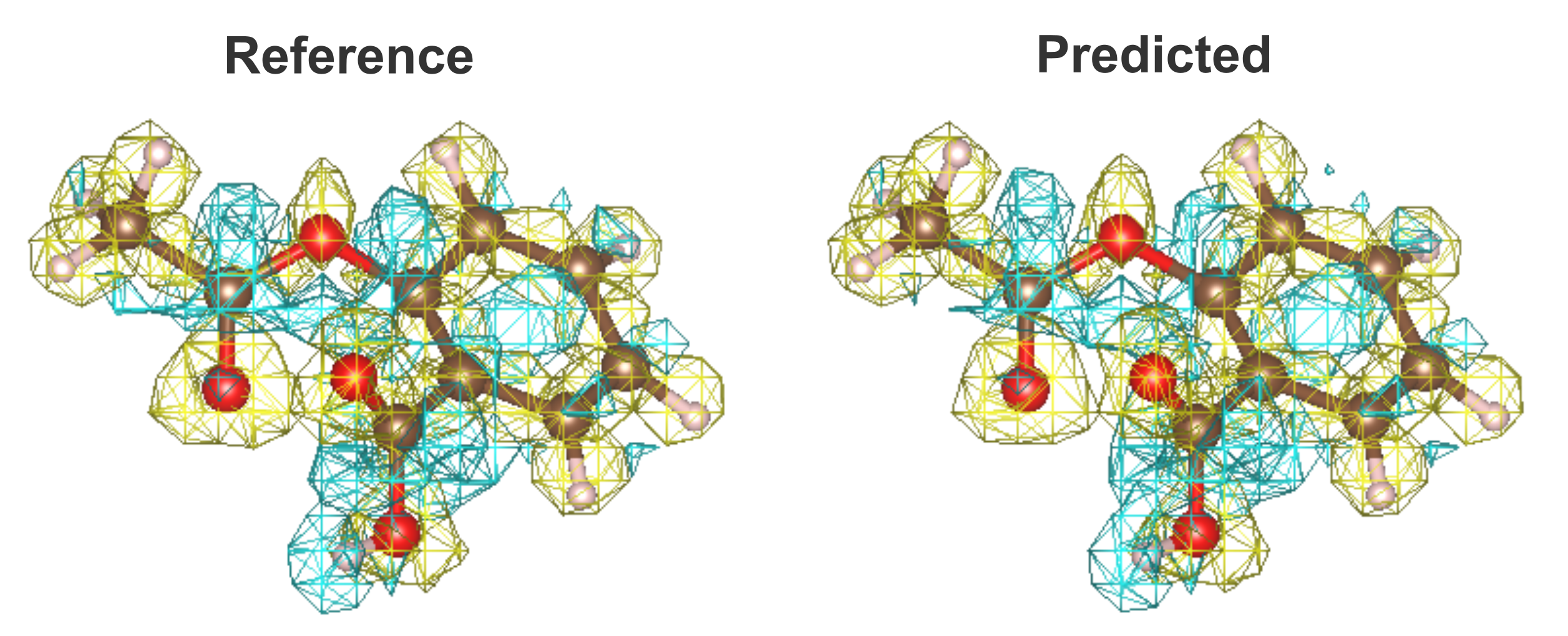}
    \caption{Comparison of the reference and predicted $\Delta\rho_g$ for a test aspirin configuration. The yellow and blue iso-surfaces represent $\Delta\rho_g$ values of $0.0095$ and $-0.0095$, respectively, which corresponds to 10\% of the maximum value of the reference $\Delta\rho_g$. Brown, white and red colored atoms denote C, H, and O species respectively.}
    \label{fig:aspirin_density}
\end{figure}

\noindent First, we discuss the results of FTA modeling on the rMD17-aspirin dataset. We train our FTA model on training set sizes ranging from 10 to 950 aspirin configurations, and test the predictions on a test set of 1000 aspirin configurations. Figure~\ref{fig:aspirin_learningcurve} shows the learning rate for force and the dipole moment of $\Delta \rho_g$, henceforth denoted as $\boldsymbol{d}^{\Delta \rho_g}$. We use two independent training approaches, one with the field loss (density and potential loss) in the objective function (cf. Eq.~\ref{eq:loss}) and the other without the field loss (i.e., with $t_\rho=0$ and $t_\phi=0$). For force prediction, both approaches have nearly identical behavior with a learning exponent of -0.41. However, for dipole moments, we observe dramatic differences between the two approaches. The model with the field loss has a learning exponent of -0.24 in $\boldsymbol{d}^{\Delta \rho_g}$, whereas the model without the field loss shows much larger dipole moment error and further a negative correlation in $\boldsymbol{d}^{\Delta \rho_g}$ errors with respect to training set size. We also compare the accuracy of ${\Delta \rho_g}$ and ${\Delta \phi_g}$ as well the first three moments of $\Delta \rho_g$, namely monopole ($m^{\Delta \rho_g}$), dipole ($\bd^{\Delta \rho_g}$),
and quadrupole moment ($\bq^{\Delta \rho_g}$), in 
Table~\ref{tab:aspirinTable} (upper half). As expected, the model with field loss attains greater accuracy in $\Delta \rho_g$ and $\Delta \phi_g$ predictions than the one without field loss. Further, the model with field loss achieves 0.007\%, 15\% and 7\% errors in  the monopole, dipole and quadrupole moments respectively, whereas the corresponding metrics for the model without field loss are two orders of magnitude higher. This comparison of the two models---with and without the field loss--- highlights the hidden degree of freedom in the PES of FTA models to predict the electron-density, that we exploit through the variational relations connecting the total energy to electron-density and electrostatic-potential. We remark that a 15\% error in dipole moment is a respectable accuracy, as the errors in the dipole moment due to the exchange-correlation approximation inherent in DFT are known to be of similar magnitude ($5-$10\%~\cite{hait2018accurate}). We also refer to Fig.~\ref{fig:aspirin_density}, where we compare the iso-surfaces of ${\Delta \rho_g}$ prediction for a test set  aspirin configuration against the reference iso-surfaces of ${\Delta \rho_{g,\text{ref}}}$ from DFT. The predicted iso-surfaces are qualitatively very close to the reference, and the model is able to reasonably capture the complex spatial structure of charge accumulation and depletion regions in the aspirin molecule~\cite{arputharaj2012topological}.  We anticipate the  seamless prediction of dipole and quadrupole moments of the electron-density in atomistics simulations to be hugely beneficial in computational chemistry and biological sciences, among other scientific domains, in providing insights into the underlying electronic-structure effects mediating relevant chemical and physical processes. For example, it is well understood that the  dipole-dipole interactions between water molecules crucially governs the liquid-vapor phase transition and  structural properties such as radial distribution functions~\cite{dang1997molecular}.  Further, the permanent dipoles of molecules play a key role in interactions with solvent and membranes that govern transport and separation mechanisms relevant for drug design and nano-filtration applications~\cite{fong2015permeability,darvishmanesh2011physicochemical}. Molecular quadrupole moments also influence inter-molecular interactions, such as between aromatic molecules like benzene that are non-polar but with highly anisotropic electron clouds~\cite{williams1993molecular}.

\noindent Next, in Tab.~\ref{tab:aspirinTable} (lower half), we examine the aspirin molecule error metrics in energy and forces for the FTA model trained with field loss and compare against those of state-of-the-art MLIPs. Remarkably, FTA outperforms all MLIPs considered in energy errors, achieving an MAE of 3.3 meV/atom  at a small training set size of 50 configurations, that drops to 2.2 meV/atom at 950 training set size.  On force MAE metrics, FTA achieves reasonable MAE errors of 42 meV/\AA~at 950 training set size, $\sim2\times$  higher than ACE MLIP, the  architecturally closest one. However, we will demonstrate in a subsequent out-of-distribution dihedral PES test on the 3BPA molecule that FTA is able to reproduce the reference PES quite closely inspite of the higher force error metrics compared to MLIPs. We remark that the force errors can be sensitive to the choice of the learning architecture. To clarify, our force learning exponent of -0.41 (Fig.~\ref{fig:aspirin_learningcurve}), while reasonable when compared to slopes for state-of-the-art MLIPs for aspirin molecule that are between $-0.4$ to $-0.7$ (cf. Fig. 2 in Supplementary of Ref.~\cite{batzner20223}), lies on the lower end of the range. The steeper slopes are obtained with the NequIP  equivariant architecture, which performs better than ACE (cf. Tab.~\ref{tab:aspirinTable}). This indicates room for further improvement of the force prediction errors by developing improved architectures for FTA models. In this context, it is pertinent to mention that  the completeness of the ACE expansion, a key result derived in Ref.~\cite{Drautz2019}, does not  apply to the field theoretic map in this work, as the derivation of the completeness argument therein is based on treating the total energy of the system as a many-body functional of relative atomic positions, a discrete set of inputs unlike field inputs.   Thus, an important future endeavor would be formulating a theoretically rigorous complete basis of polynomial features to  approximate a field to energy map   while preserving the necessary symmetries. 


\begin{table}[h!]
\centering
\resizebox{0.85\textwidth}{!}{
\begin{tabular}{l|cccccccc}
\hline
\multicolumn{6}{c}{\textbf{FTA} ($N_{\mathrm{train}} = 950$)} \\
\hline
Aspirin molecule & $e_1$($\Delta \rho_{g}$) & $e_1$($\Delta \phi_{g}$) & MAE($m^{\Delta \rho_g}$/$N_e$) & $e_2$($\boldsymbol{d}^{\Delta \rho_g}$) & $e_2$($\boldsymbol{q}^{\Delta \rho_g}$)\\
\hline
w/ field loss &  $1.5\times10^{-3}$ & $4.2\times10^{-2}$ &  $6.8\times10^{-5}$  &  0.15 & 0.07 \\
\hline
w/o field loss & $8.6\times10^{-2}$ & $9.2\times10^{-2}$ & $5.9\times10^{-2}$& 10.2 & 5.2 \\ 
\hline
\end{tabular}
}

\vspace{0.1em} 

\resizebox{\textwidth}{!}{
\begin{tabular}{l|cccccccc|cccc}
\hline
Aspirin& \multicolumn{8}{c}{$N_{\mathrm{train}} = 950$} & \multicolumn{4}{c}{$N_{\mathrm{train}} = 50$} \\
\cline{2-13}
molecule& MACE & NequIP & ACE & sGDML & GAP & ANI & GRACE & \textbf{FTA} & ACE & NequIP & GRACE & \textbf{FTA}\\
\hline
Energy & 2.2 & 2.3 & 6.1 & 7.2 & 17.7 & 16.6 & 1.7 & 2.2 & 26.2 & 19.5 & 11.7 & 3.3\\
Forces & 6.6 & 8.2 & 17.9 & 31.8 & 44.9 & 40.6 & 6.1 & 42.5 &63.8 & 52.0 & 38.2 & 134.6\\
\hline
\end{tabular}
}
\caption{Benchmark results for rMD17-aspirin. 
Top table:  $\Delta \rho_g$ and $\Delta \phi_g$ errors for FTA with and without field loss in training. $m^{\Delta \rho_g}$,  $\bd^{\Delta \rho_g}$, $\bq^{\Delta \rho_g}$ denote the monopole, dipole, and quadrupole moments respectively of $\Delta \rho_g$. The $e_1$ and $e_2$ error metrics are defined as follows: $e_1(f)=\sum_i \left(\frac{ \norm{f^i_{\text{pred}}-f^i_{\text{ref}}}}{\norm{f^i_{\text{ref}}}}\right)/M$, and $e_2(f)=\frac{\sum_i \norm{f^i_{\text{pred}}-f^i_{\text{ref}}}}{\sum_i \norm{f^i_{\text{ref}}}}$, where $M$ denotes the number of test systems. In $e_1$ and $e_2$, we use Coulomb $J$ norm for $\Delta \rho_g$, $L_2$ norm for $\Delta \phi_g$ and  $\boldsymbol{d}^{\Delta \rho_g}$, and Frobenius norm for $\boldsymbol{q}^{\Delta \rho_g}$.
Bottom table: Comparison of energy and force MAEs of FTA (with field loss) against  MLIPs. The energy units are in meV/atom and force units are in meV/\AA. The benchmark results for  MLIPs are taken from the following Refs.~\cite{3BPA,bochkarev2024graph,batatia2022mace}.}
\label{tab:aspirinTable}
\end{table}

\begin{table}[h!]
\centering
\resizebox{0.9\textwidth}{!}{
\begin{tabular}{l|cccccccc}
\hline
\multicolumn{6}{c}{\textbf{FTA} ($N_{\mathrm{train}} = 500$ on 300 K data)} \\
\hline
3BPA molecule & $e_1$($\Delta \rho_{g}$) & $e_1$($\Delta \phi_{g}$) & MAE($m^{\Delta \rho_g}$/$N_e$) & $e_2$($\boldsymbol{d}^{\Delta \rho_g}$) & $e_2$($\boldsymbol{q}^{\Delta \rho_g}$)\\
\hline
1200 K   & $1.9\times 10^{-3}$ &    $4.1\times 10^{-2}$ & $5.4\times 10^{-5}$  & 0.25 & 0.24 \\
\hline
$\beta=120^{\circ}$ dihedral slice   & $7.5\times 10^{-4}$ & $3.4\times 10^{-2}$   & $2\times 10^{-5}$  & 0.17 & 0.08\\
\hline
\end{tabular}
}

\vspace{0.1em} 

\resizebox{0.95\textwidth}{!}{
\begin{tabular}{l|cccccccc}
\hline
3BPA  molecule & \multicolumn{7}{c}{$N_{\mathrm{train}} = 500$ on 300 K data}  \\
\cline{2-9}
& MACE & NequIP & ACE & sGDML & GAP & ANI & BOTNet & \textbf{FTA} \\
\hline
1200 K Energy & 29.8 & 40.8 & 85.3 & 774.5 & 166.8 & 76.8  & 39.1 &  4.4 \\
1200 K Forces & 62.0 & 86.4 & 187.0 & 711.1 & 305.5 & 129.6 & 81.9 &  307.2 \\
All dihedral slices Energy & 7.8 & 23.2 & - & - &  - & -  & 16.3 &  1.3 \\
All dihedral slices Forces & 16.5 & 23.1 & - & - & - & -  &  20.0 & 60.1    \\
\hline
\end{tabular}
}
\caption{3BPA benchmark results on out-of-distribution tests: 1200 K MD trajectory samples and dihedral slices. 
Top table: $\Delta \rho_g$ and $\Delta \phi_g$ errors for FTA. We refer to Table~\ref{tab:aspirinTable}'s caption for other notations and definition of the $e_1$ and $e_2$ error metrics.
Bottom table: comparison of energy and force RMSE of FTA against  MLIPs.  The energy units are in meV/atom and force units are in meV/\AA. The FTA model is trained with field loss terms. The benchmark results for  MLIPs are taken from the following Refs.~\cite{3BPA,bochkarev2024graph,batatia2022mace}. The dashed entries in the lower table denote unavailable data.}
\label{tab:3BPATable}
\end{table}

\begin{figure}[htbp]
  \centering
  \includegraphics[scale=0.16]{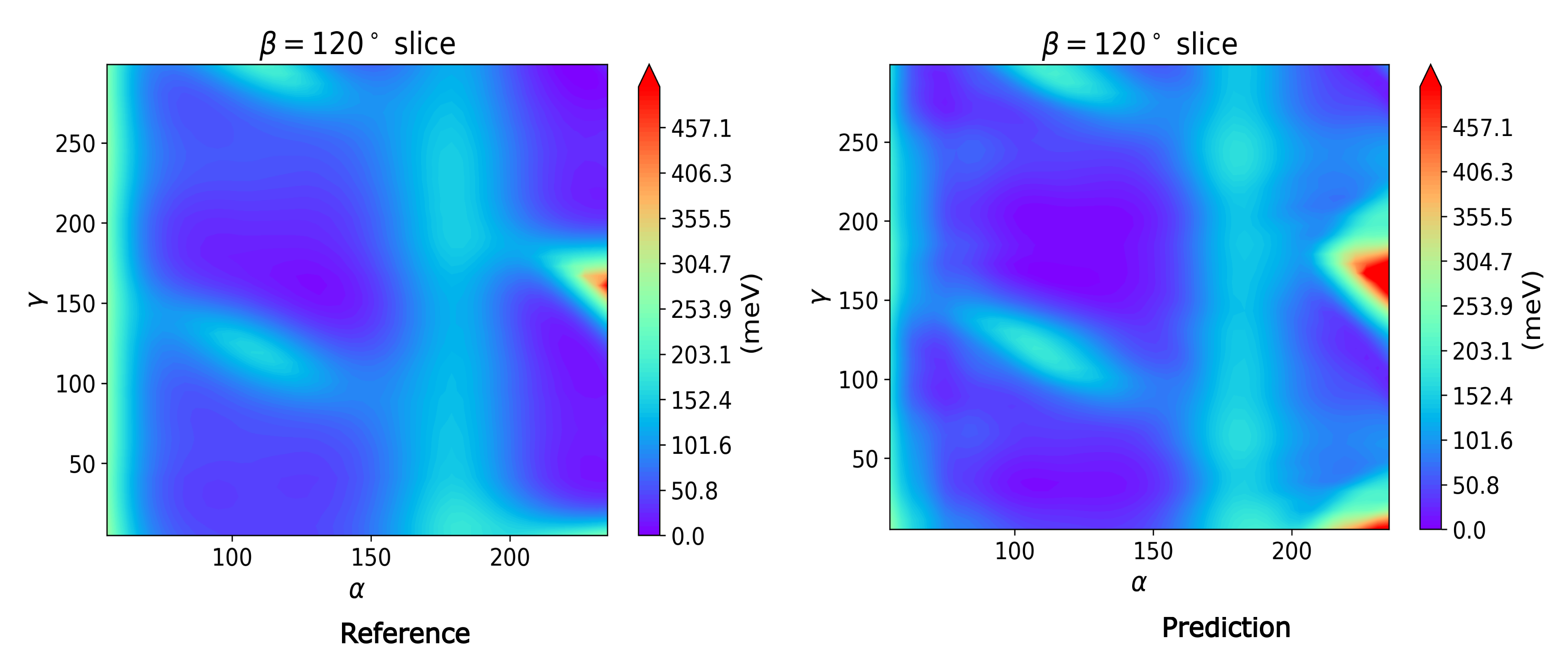}
  \caption{Comparison of the potential energy surface of 3BPA for the $\beta=120^{\circ}$ dihedral slice.}
  \label{fig:3BPASlicePES}
\end{figure}

\begin{figure}[htbp]
  \centering
  \includegraphics[scale=0.16]{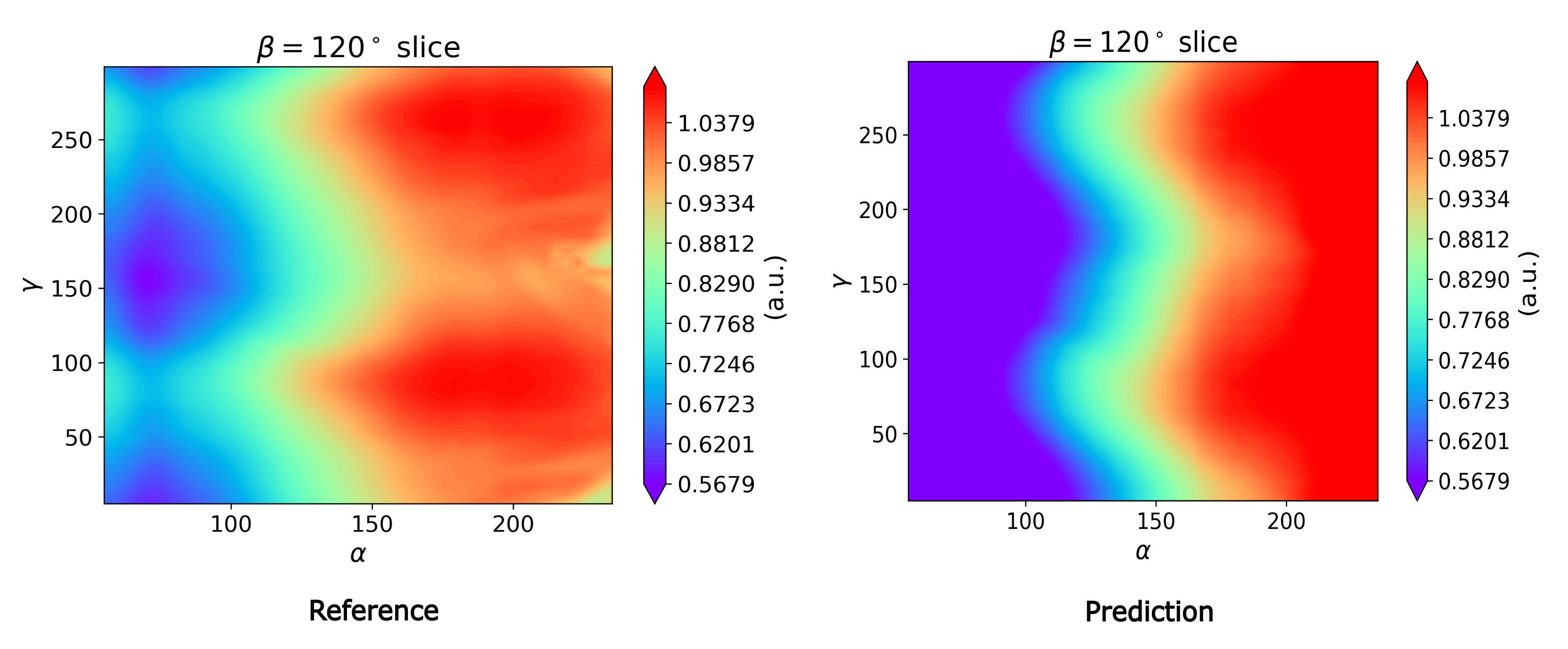}
  \caption{Comparison of the dipole moment magnitude ($\abs{\boldsymbol{d}^{\Delta \rho_g}}$) of 3BPA for the $\beta=120^{\circ}$ dihedral slice.}
  \label{fig:3BPASliceDipoleDensity}
\end{figure}

\begin{figure}[htbp]
  \centering
  \includegraphics[width=0.95\textwidth]{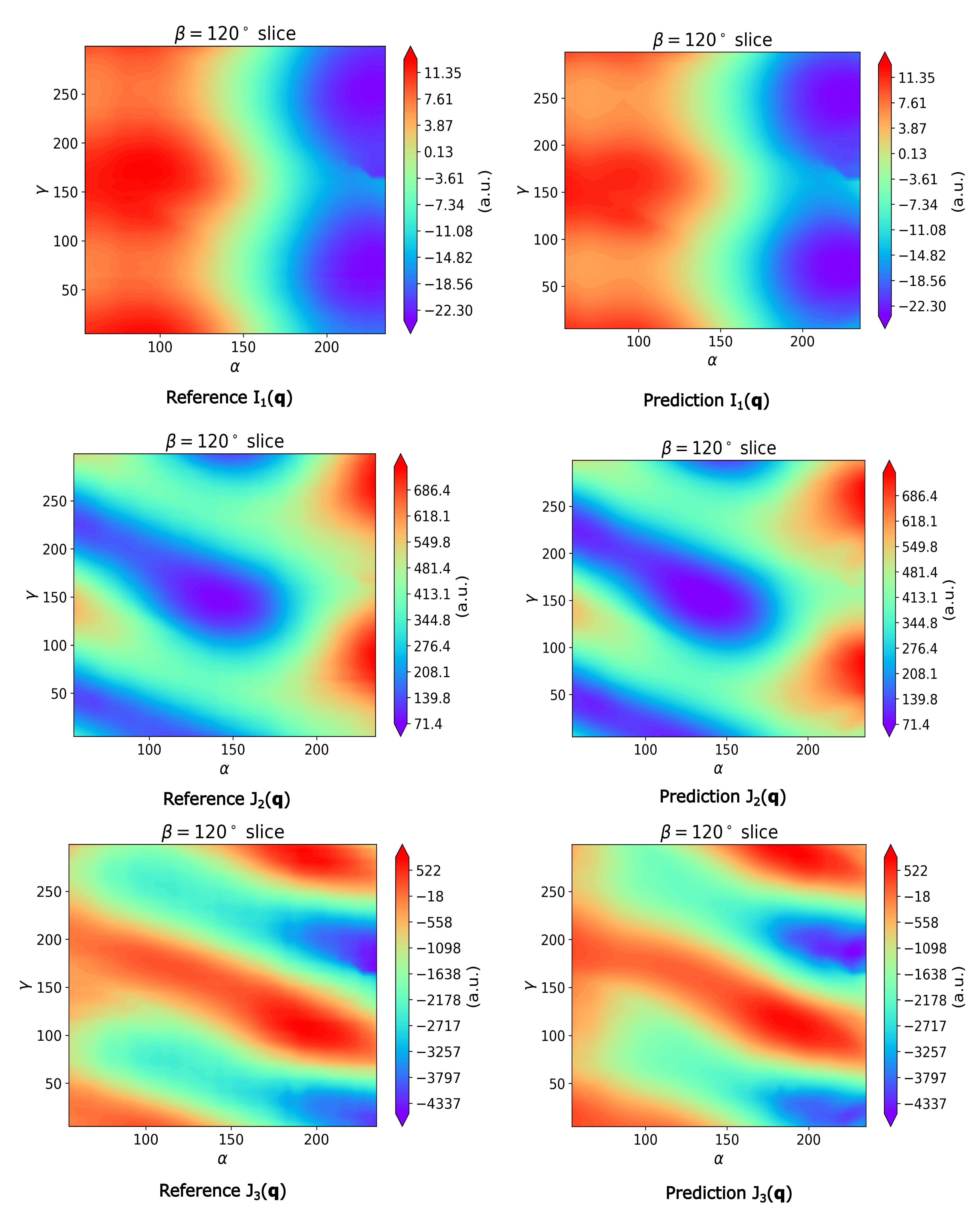}
  \caption{Comparison of the three  invariants of the quadrupole moment tensor ($\boldsymbol{q}^{\Delta \rho_g}$) of 3BPA for the $\beta=120^{\circ}$ dihedral slice. The expressions for the various invariants are given by $I_1(\boldsymbol{q}^{\Delta \rho_g}) = \textrm{Tr}(\boldsymbol{q}^{\Delta \rho_g})$, $J_2(\boldsymbol{q}^{\Delta \rho_g}) = \frac{1}{2} \norm{\boldsymbol{q}^{\Delta \rho_g} - \frac{1}{3} I_1 (\boldsymbol{q}^{\Delta \rho_g})\mathbf{I}}_F^2$, and  $J_3(\boldsymbol{q}^{\Delta \rho_g}) = \textrm{det}(\boldsymbol{q}^{\Delta \rho_g} - \frac{1}{3} I_1 (\boldsymbol{q}^{\Delta \rho_g})\mathbf{I})$.}
  \label{fig:3BPASliceQuadrupoleDensity}
\end{figure}

\noindent To demonstrate the out-of-distribution performance of the FTA models, we apply it to the 3-(benzyloxy)pyridin-2-
amine (3BPA) dataset~\cite{3BPA}. 3BPA has three consecutive rotatable bonds which leads to
complex dihedral potential energy surface (PES) with
several local minima, and hence, constitutes a stringent out-of-distribution test. The training set consists of 500  configurations of the 3BPA from molecular dynamics (MD) trajectories at 300K. The testing, in terms of energy, force, and field predictions is done using two out-of-distribution sets: one with configurations from 1200K MD trajectories configurations and the other with dihedral rotations. The test set structures are farther away from the training set compared to the rMD17-aspirin benchmark. Particularly, doing well on the dihedral PES test demands a smooth extrapolation of the fitted model as they involve out-of-equilibrium geometries. Table~\ref{tab:3BPATable} shows the relevant results, comparing the energy and force RMSE errors of FTA against widely used MLIPs. First, we observe that the FTA model outperforms all the MLIPs considered in energy predictions, achieving an accuracy of a few 1--5 meV/atom on both 1200K configurations and dihedral slices.  In comparison, the best performing MLIP, MACE achieves 8--30 meV/atom energy errors. We remark that MACE has a more expressive learning architecture compared to the ACE architecture used in the FTA model. Next, we observe that the force errors of FTA for both the 1200K and dihedral test sets is about 1.6$\times$ higher than the ACE model, which is architecturally closest to the FTA model.  Notwithstanding the higher force RMSE errors, in Fig.~\ref{fig:3BPASlicePES}, we demonstrate that FTA closely reproduces the reference DFT dihedral PES for the $\beta=120^{\circ}$ slice (slice that is farthest away from training points~\cite{3BPA}), with good qualitative and quantitative agreement of the peaks and the valleys against the reference. The above results on energy and forces on both the rMD17-aspirin and 3BPA benchmarks suggest that the FTA approach is competitive against state-of-the-art MLIPs. Finally, we discuss the field predictions for 3BPA dataset. In the upper half of Tab.~\ref{tab:3BPATable}, we demonstrate that FTA attains good error metrics on $\Delta \rho_g$ and $\Delta \phi_g$, with the $e_1$ metrics comparable to the rMD17-aspirin case. Notably, our   $\boldsymbol{d}^{\Delta \rho_g}$ and   $\boldsymbol{q}^{\Delta \rho_g}$ relative error metrics are 17--25\% and 8--24\% respectively, even in the out-of-distribution tests. To further probe the quality of the $\boldsymbol{d}^{\Delta \rho_g}$ and   $\boldsymbol{q}^{\Delta \rho_g}$ predictions, in Fig.~\ref{fig:3BPASliceDipoleDensity} and Fig.~\ref{fig:3BPASliceQuadrupoleDensity}, we plot the  magnitude of $\boldsymbol{d}^{\Delta \rho_g}$ and invariants of $\boldsymbol{q}^{\Delta \rho_g}$  tensor ($I_1$, $J_2$, and $J_3$)  as a function of the $\alpha$ and $\gamma$ dihedral angles on the $\beta=120^{\circ}$ slice and compare with the reference DFT values.  Physically, the  $I_1$ invariant of $\boldsymbol{q}^{\Delta \rho_g}$, which is its trace, measures the spatial spread of $\Delta \rho_g$ in an isotropic sense. We also considered the anisotropic effects in the dihedral slice through the $J_2$ and $J_3$ invariants that are functions of the traceless quadrupole moment tensor ($\boldsymbol{q}^{\Delta \rho_g} - \frac{1}{3} I_1 (\boldsymbol{q}^{\Delta \rho_g})\mathbf{I}$). We remark that  $J_2$ and $J_3$ invariants are relevant for electrostatic interactions as in the multi-pole expansion for the electrostatic potential of a charge distribution, the $\frac{1}{R^3}$ term is dependent on the traceless quadrupole moment of the charge distribution.   As evident from  Fig.~\ref{fig:3BPASliceDipoleDensity} and Fig.~\ref{fig:3BPASliceQuadrupoleDensity}, we observe a good qualitative agreement with the reference DFT data in the $\boldsymbol{d}^{\Delta \rho_g}$ magnitude, and excellent agreement in the $\boldsymbol{q}^{\Delta \rho_g}$ invariants. It is interesting to note the complex structure of the $\boldsymbol{d}^{\Delta \rho_g}$ and $\boldsymbol{q}^{\Delta \rho_g}$ slice plots, which quite likely  contain important qualitative information about charge re-distribution due to the steric hindrance caused by rotating bonds in 3BPA. Our results demonstrate that  FTA theory is able to reliably predict the electronic-structure underlying the molecular PES, both in interpolative and extrapolative regimes.  We believe that the variational origin of our ${\Delta \rho_g}$ prediction (cf. Eq.~\ref{eq:deltaSlopeDeltaRho1}), as opposed to learning a direct map from  $(\vaux,\bsmear)$  to ${\Delta \rho_g}$ (untethered to the PES) plays a key role in this regard. 

%% file: discussion.tex
\section{Discussion} \label{sec:discussion}
In this work, we introduce field theoretic atomistics (FTA) as a theoretically rigorous and practically useful large-scale surrogate to DFT or any electronic structure method, in general. In FTA, we reformulate the Hohenberg-Kohn (HK) energy map in terms of auxiliary fields ($\vaux(\br)$ and $\bsmear(\br)$) to lend it thermodynamic and variational consistency. FTA uses the HK theorem to define the groundstate density and atomic forces as variation of the energy with the input auxiliary fields. This field theoretic and variational aspect makes FTA an electronic structure method, unlike interatomic potentials, where the structure to energy map omits all electronic information. Our machine-learned FTA models exhibit good accuracy for energies, forces, and electron densities, including for out-of-distribution benchmarks. Particularly noteworthy is their ability to reproduce accurate dipole and quadrupole moments from the predicted densities.

\noindent The explicit treatment of electron density in FTA enables access to phenomena such as electron–phonon coupling in molecular and condensed-matter systems~\cite{hauf2019phonon, Giustino2017}. Further, FTA, by virtue of its field theoretic formulation, can naturally incorporate any externally applied electric field. This allows it to seamlessly couple with the solvation models used in electrochemical and biochemistry  applications~\cite{Mennucci2012polarizable, Kleinjung2014design}. 
While FTA introduces a new paradigm for atomistic modeling, achieving chemical accuracy across diverse chemical environments demands further development. One key direction is to enhance FTA's generalization to systems with significant long-range interactions through improved machine-learning architectures and nonlocal descriptors. Another promising avenue is to extend the FTA formulation to variable number of electrons (grand-canonical ensemble), thereby enabling large-scale atomistics in open systems with a fixed chemical potential. Overall, we expect the combination of electronic-structure foundation and scalability in FTA to open new avenues and provide deeper insights across computational chemistry and materials science.


%% file: method.tex
\section{Methods} \label{sec:method}
\subsection*{DFT calculations}
The  DFT computations of the ground-state energy, forces and electron-density in this work were performed using the PySCF package~\cite{sun2020recent,paszke2019pytorch}. For the rMD17-aspirin calculations we employed PBE exchange-correlation functional  and the def2-SVP basis set. In case of the 3BPA, we employed $\omega\text{B97X}$ exchange-correlation functional  and the 6-31G(d) basis set. We use a refined PySCF quadrature grid (level 8) to evaluate the ground-state density, required for training our FTA models.

\subsection*{Description of the datasets}
For all the datasets used in this work, we have recomputed the reference DFT data on ground-state energy, forces, and electron-density using the PySCF software.
\paragraph{Aspirin molecule from rMD17 dataset:}
The revised MD17 (rMD17) dataset consists of MD trajectory samples of small organic
molecules~\cite{rmd17}.  In particular, the dataset has five train-test splits for each molecule, with each split containing 1000 train and 1000 test configurations sampled randomly from the AIMD trajectory at 500 K.  In this work, we consider only the aspirin molecule (2-(acetyloxy)benzoic acid), the largest system in the dataset. The aspirin molecule consists of 21 atoms and (C,H,O) chemical species.
\paragraph{3BPA dataset:}
The 3BPA dataset contains train-test splits of a flexible drug-like organic molecule (3-(benzyloxy)pyridin-2-amine)~\cite{3BPA,3BPAMoleculeOriginal}. The 3BPA molecule consists of 27 atoms and (C,H,N,O) chemical species. There are 500 training
snapshots sampled randomly from the AIMD trajectory at 300 K. The test sets consist of AIMD trajectory snapshots at 600 K and 1200 K, and a torsional PES test set sampled along dihedral rotations of the three rotatable bonds. In the torsional PES test set, three 2D dihedral slices are considered for $\beta=120^{\circ}$,  $\beta=150^{\circ}$, and  $\beta=180^{\circ}$. In particular, $\beta=120^{\circ}$ slice has very few nearby training set points from the 300 K trajectory~\cite{3BPA}.  In this work, we study the 1200 K MD trajectory samples and torsional PES test sets.

\subsection*{Equivariant machine learning architecture}
We recall that our model is expressed as a separable form (cf. Eq.~\ref{eq:separableForm}), where one branch of the model is  $\DTE_{I,A}[\Delta\bx_I, \Delta \by_I]$ and the other branch is $\DTE_{I,B}[ \Delta\bx_I, \Delta \by_I, \bx_I,  \by_I]$. Below we discuss our learning architecture to model both the branches while ensuring the well-separated limit is satisfied for the product of the branches $\DTE_I$,  an exact condition (cf. Sec.~\ref{sec:results}). Broadly, for our learning architecture, we choose a combination of atomic cluster expansion (ACE)~\cite{Drautz2019} and deep neural networks (NN) as depicted in Fig.~\ref{fig:schematicFTA}.  In the first step, we employ the ACE expansion to learn rotationally invariant features from our  descriptors ($\Delta\bx_I, \Delta \by_I, \bx_I,  \by_I$). The descriptors are obtained by sampling the input fields ($\vaux$,$\bsmear$) using atom centered probe functions that are represented as separable products of radial functions and real spherical harmonics (cf. Eq.~\ref{eq:descriptors}). As a result, our descriptors at each atomic site $I$,  generically denoted by $\boldsymbol{D}_I$, transform equivariantly with respect to the input fields. The ACE expansion for a rotational invariant scalar $P$ at an atomic site $I$ as a function of  $\boldsymbol{D}_I$ is defined as the following polynomial expansion~\cite{Drautz2019}
\begin{align}\label{eq:ACEExpansion}
P(\boldsymbol{D}_I;K) = \sum_{v=1}^{v=K}  \sum_{nl} c_{nl}^{(v)} B_{Inl}^{(v)}\,,
\end{align}
where the expansion is truncated at the $K^{\text{th}}$ correlation order, $c_{nl}^{(v)}$ are learnable parameters, and  $B_{Inl}^{(v)}$ is the hierarchial many-body ACE basis. The $B_{Inl}^{(v)}$ tensors in the ACE expansion are constructed from $v^{\text{th}}$ order rotationally invariant products of the $D_{Inlm}$ tensor, using properties of product of spherical harmonic tensors. Thus, the ACE expansion is able to learn  higher-order features from $D_{Inlm}$. 

\noindent Since ACE is a linear expansion in a many-body basis, next we leverage NN  for learning more expressive nonlinear functional forms on the rotationally invariant ACE outputs. Specifically, we use two NN layers: a $f_A^{(\mathrm{NN})}$ for the $\DTE_{I,A}$ branch and a $f_B^{(\mathrm{NN})}$  for the $\DTE_{I,B}$ branch. Further, we use multiple ACE channels $N_c$ for the $\Delta$-field descriptors ($\Delta\bx_I, \Delta \by_I$) and use one ACE channel for the total-field descriptors ($\bx_I, \by_I$), and concatenate the scalar outputs $P$ in the following manner to form input feature vectors  into the NN layer
\begin{align}\label{eq:NNarch}
   \boldsymbol{Q}^{(N_c)}_{\boldsymbol{D}_I}=&\left[ P^{(1)}(\boldsymbol{D}_I)\,,\cdots,\, P^{(N_c)}(\boldsymbol{D}_I)\right]\,,\notag\\
\DTE_{I,A} = & f_A^{(\mathrm{NN})}\left(\left[\boldsymbol{Q}_{\Delta \bx_I}^{(N_c)} ,\, \boldsymbol{Q}_{\Delta \by_I}^{(N_c)} \right]\right),\notag\\
\DTE_{I,B} = & f_B^{(\mathrm{NN})}\left(\left[\boldsymbol{Q}_{\Delta \bx_I}^{(N_c)} ,\, \boldsymbol{Q}_{\Delta \by_I}^{(N_c)},\,\boldsymbol{Q}_{\bx_I}^{(1)} ,\, \boldsymbol{Q}_{\by_I}^{(1)}) \right]\right)\,,
\end{align}
where $\boldsymbol{Q}_{\Delta \bx_I}^{(N_c)}$ denotes $\left[ P^{(1)}({\Delta \bx_I})\,,\cdots,\, P^{(N_c)}({\Delta \bx_I})\right]$, and accordingly for the other field descriptor types. We note that $f_A^{(\mathrm{NN})}$ only takes $\Delta$-field descriptors while $f_B^{(\mathrm{NN})}$ takes both  $\Delta$- and total-field descriptors. Further, we set the bias parameters in $f_A^{(\mathrm{NN})}$ to be zero and choose activation functions that are zero valued at zero input. The above construction choices in conjunction with  $\boldsymbol{Q}_{\Delta \bx_I}^{(N_c)}$ and $\boldsymbol{Q}_{\Delta \by_I}^{(N_c)}$ being zero when $\Delta \bx_I$ and $\Delta \by_I$ are zero (property of the ACE expansion) ensures that  $\DTE_{I,A}$ tends to zero at the well-separated/isolated atoms limit. Thus, we have now a learning architecture that is exact in atomization energy, forces, $\Delta \rho_g$ and $\Delta \phi_g$ in the well-separated/isolated atoms limit. In the above, we remark that the disparity in number of channels between $\Delta$-field and total-field ACE channels is based on our empirical observations that higher number of  $\Delta$-field ACE channels attain lower training loss on $\Delta \rho_g$ and $\Delta \phi_g$.


\subsection*{Model and training parameters}
We begin with discussing the choices of the weights in the objective function, then the probe function parameters, and finally the parameters in the learning architecture. 
In both the rMD17-aspirin and 3BPA dataset training, we assign the following weights in the loss function (cf. Eq.~\ref{eq:loss}): $t_E=1\times 10^{7}$, $t_F=3\times 10^{10}$, $t_\rho=1.8\times 10^{7}$, and $t_\phi=3\times 10^{3}$ corresponding to the energy, force, density, and potential loss terms. Next, the parameters for the atomic site  centered  probe functions to compute the rotationally equivariant descriptors ($\bx_I$, $\by_I$, $\Delta \bx_I$, $\Delta \by_I$) of the input fields ($\vaux$ and $\bsmear$) are set as follows. We use a combination of short-ranged orthogonal Bessel (SR-OB) functions and long-ranged damped-multipole
(LR-DM) functions for the radial portion of our probe functions, and $l^{\text{max}}=3$ for the angular portion. We use a total of $22\, (20+2)$ and $17\, (15+2)$ radial probe functions for $\vaux$ and $\bsmear$ respectively, with radial cut-offs of 4--6.5 Bohr for SR-OB functions and around 10 Bohr for LR-DM functions. We refer to Sec. S2 of the SI for details of their construction. The same probe function choices are used for both rMD17-aspirin and 3BPA training.  Subsequently, for the ACE expansion, that uses the descriptors as inputs,  we set the correlation order to be $K=4$  and $K=2$ for the $\DTE_{I,A}$ and $\DTE_{I,B}$ respectively. We motivate this choice from the observation that $\Delta \bx_I$ and $\Delta \by_I$ have a relatively smaller variance in their component values across the angular momentum channels compared to ${\bx_I}$ and ${\by_I}$ which have large values for the $l=0$ components due to the dominant spherically symmetric self-atom portion of the input fields. Thus $\Delta \bx_I$ and $\Delta \by_I$ should be more amenable to learning higher order correlations in the ACE expansion that are expected to be smaller in magnitude compared to the lower order terms.  We also sparsify the ACE expansion through  limiting the maximum angular momentum and radial basis index for the higher correlation orders ($K\ge3$) to avoid over-parametrization. In particular, we use $(n^{\text{max}}=10,\,l^{\text{max}}=3)$ for $K=3$ and $(n^{\text{max}}=4,\,l^{\text{max}}=2)$ for $K=4$. Further, we use 40 ACE channels for $\Delta$-field descriptors ($N_c=40$ in Eq.~\ref{eq:NNarch}) and 1 ACE channel for total-field descriptors. The output of the ACE channels is transferred into the NN layer. For the NN layer, in both $\DTE_{I,A}$ and $\DTE_{I,B}$, we use feed-forward neural network consisting of two hidden layers and SiLU activation functions. The number of neurons in each layer is coarsened by a factor 0.75 starting from the input layer. The input layer size depends on the number of ACE channels. The feed-forward networks are implemented via the Pytorch software package~\cite{paszke2019pytorch}.

\noindent The  ACE parameters in all the channels, and weights and biases in the NN layer are optimized together using the AdamW optimizer~\cite{loshchilov2017fixing}, with a weight decay of $5\times 10^{-5}$, a mini-batch size of 16, and an initial learning rate of $6\times 10^{-3}$ that is adaptively damped by a factor 0.8 if the valid loss does not decrease for 600 epochs or more. We continue the optimization to a maximum of 150,000 epochs or if the validation loss over 1000 consecutive epochs does not reduce, whichever condition is satisfied first.

%% file: SI.tex
\section{Bijective Maps and Variational Relations} \label{SIsec:maps}
In this section, we derive the bijective maps and the variational relations presented in the main manuscript. We start with the saddle point reformulation of DFT
\begin{equation}\label{SIeq:DFTSaddle}
    \TE[\vaux, \bsmear, N_e] = \min_{\rho\in A_N} \max_{\phi} \left(\Ftilde[\rho] + \int \rho(\br) \vaux(\br)\dr + P[\phi, \rho, \bsmear]\right), \, \text{s.t.}\, \int \rho(\br)\dr = N_e\,,
\end{equation}
\subsection{Relation between $(\rho_g, \mu_g)$ and $(\vaux, N_e)$} \label{SIsec:rhov}
Let us denote $\bar{P}[\rho,\bsmear] = \max_{\phi} P[\phi, \rho, \bsmear]$. Also, let us introduce a Lagrange multiplier $\mu$ for the constraint $\int \rho(\br)\dr = N_e$. As will be shown later, the groundstate $\mu$ is the chemical potential of the system. Thus, using Eq.~\ref{SIeq:DFTSaddle}, the groundstate density ($\rho_g$) can be obtained by solving
\begin{equation} \label{SIeq:LRho}
\min_{\rho}\left(\Ftilde[\rho] + \int \rho(\br)\vaux(\br)\dr + \bar{P}[\rho,\bsmear]-\mu\left(\int\rho(\br)\dr-N_e\right)\right) = 0\,,   
\end{equation}
which is equivalent to solving
\begin{equation}\label{SIeq:dLRho}
\frac{\delta (\Ftilde[\rho] + \bar{P}[\rho,\bsmear])}{\delta \rho(\br)} + \vaux(\br) = \mu\,.   
\end{equation}
In the above, $\mu$ has to be updated until $\int \rho(\br)\dr = N_e$. We now prove the bijective map between $(\rho_g,\mu_g)$ and $(\vaux, N_e)$. The proof is by contradiction, akin to the proof of the original HK theorem. First of all, two different $N_e$'s cannot lead to the same $\rho_g$, as the same density cannot integrate to different number of electrons. Thus, to show the bijectivity between $(\rho_g,\mu_g)$ and $(\vaux, N_e)$, we need to show that, for a given $N_e$, two different $\vaux$'s cannot result in the same $\rho_g$ and $\mu_g$. To prove the above, let us assume the contrary: that, for a fixed $N_e$, two different potentials---$\vaux^{(1)}(\br)$ and $\vaux^{(2)}(\br)$---yield the same $\rho_g$ and $\mu_g$. 
Using Eq.~\ref{SIeq:dLRho}, we have
\begin{equation}
\left.\frac{\delta (\Ftilde[\rho] + \bar{P}[\rho,\bsmear])}{\delta \rho(\br)}\right|_{\rho_g} + \vaux^{(1)}(\br) = \mu_g
\end{equation}
\begin{equation}
\left.\frac{\delta (\Ftilde[\rho] + \bar{P}[\rho,\bsmear])}{\delta \rho(\br)}\right|_{\rho_g} + \vaux^{(2)}(\br) = \mu_g
\end{equation}
Taking the difference of the above two equations leads to $\vaux^{(1)}(\br) - \vaux^{(2)}(\br)=0$, which contradicts our original assumption. This proves the bijectivity between $(\rho_g,\mu_g)$ and $(\vaux, N_e)$

We now prove that $\frac{\delta \TE[\vaux,\bsmear,N_e]}{\delta\vaux(\br)} = \rho_g(\br)$. At groundstate, $\TE[\vaux,\bsmear, N_e] = \Ftilde[\rho_g] + \bar{P}[\rho_g,\bsmear] + \int\rho_g(\br)\vaux(\br)$. Thus,
\begin{equation}
\begin{split}
\frac{\delta \TE[\vaux,\bsmear,N_e]}{\delta \vaux(\br)} &= \mathlarger{\int} \left(\frac{\delta (\Ftilde[\rho] + \bar{P}[\rho,\bsmear])}{\delta \rho_g(\br')} + \vaux(\br')\right)\frac{\delta \rho_g(\br')}{\delta \vaux(\br)}\dr' + \rho_g(\br)\\
&=\mu_g\int \frac{\delta \rho_g(\br')}{\delta \vaux(\br)}\dr' + \rho_g(\br)\\
&= \rho_g(\br)\,, 
\end{split}
\end{equation}
where in the second line we used Eq.~\ref{SIeq:dLRho}. In the third line, we used the fact that $\int\rho_g(\br')\dr' = N_e$, which results in $\int \frac{\delta \rho_g(\br')}{\delta \vaux(\br)}\dr' = 0 $. 

We can, similarly, prove $\frac{\partial \TE[\vaux, \bsmear, N_e]}{\partial N_e} = \mu_g$. We have
\begin{equation}
\begin{split}
    \frac{\partial \TE[\vaux,\bsmear,N_e]}{\partial N_e} &= \mathlarger{\int} \left(\frac{\delta (\Ftilde[\rho] + \bar{P}[\rho,\bsmear])}{\delta \rho_g(\br')} + \vaux(\br')\right)\frac{\partial \rho_g(\br')}{\partial N_e}\dr' \\
    & = \mu_g \int \frac{\partial \rho_g(\br')}{\partial N_e}\dr'\\
    &= \mu_g\,,
\end{split}
\end{equation}
where in the second line we used Eq.~\ref{SIeq:dLRho}. In the third line, we used the fact that $\int\rho_g(\br')\dr' = N_e$, which results in $\int \frac{\partial \rho_g(\br')}{\partial N_e}\dr' = 1 $. 

\subsection{Relation between $\phi_g$ and $\bsmear$} \label{SIsec:phib}
We recall that $P[\phi, \rho, \bsmear] = -\frac{1}{8\pi}\int |\nabla\phi(\br)|^2\dr + \int\left(\rho(\br)+\bsmear(\br)\right)\phi(\br)\dr$. Let us define $G[\phi, \vaux, N_e]$ as
\begin{equation}
G[\phi, \vaux, N_e] = \min_\rho \left(\Ftilde[\rho] + \int (\vaux(\br) +  \phi(\br))\rho(\br)\dr\right)\,,\quad\text{s.t.~} \int \rho(\br)\dr = N_e\,. 
\end{equation}
Then, using Eq.~\ref{SIeq:DFTSaddle}, the groundstate electrostatic potential, $\phi_g$, can be found by solving
\begin{equation} \label{SIeq:LPhi}
    \max_{\phi} \left(-\frac{1}{8\pi}\int|\nabla\phi(\br)|^2\dr+\int \bsmear(\br) \phi(\br)\dr + G[\phi,\vaux,N_e]\right)\,,
\end{equation}
which is equivalent to 
\begin{equation}\label{SIeq:dLPhi}
    \frac{1}{4\pi}\nabla^2\phi_g(\br) + \bsmear(\br) + \frac{\delta G[\phi_g,\vaux, N_e]}{\delta \phi_g(\br)}= 0\,. 
\end{equation}
To prove the bijectivity between $\phi_g$ and $\bsmear$, we use similar arguments of proof by contradiction used in Sec.~\ref{SIsec:rhov}. We assume there are two different nuclear charges, $\bsmear^{(1)}$ and $\bsmear^{(2)}$, which result in the same groundstate electrostatic potential ($\phi_g$). Thus, both $\bsmear^{(1)}$ and $\bsmear^{(2)}$ should satisfy Eq.~\ref{SIeq:dLPhi} at $\phi_g$. Taking the difference of those two equations will lead to the condition: $\bsmear^{(1)}(\br)-\bsmear^{(2)}(\br)=0$, which contradicts our original assumption.      

Using Eq.~\ref{SIeq:LPhi}, we can rewrite $\TE$ as
\begin{equation} 
    \TE[\vaux,\bsmear, N_e] = -\frac{1}{8\pi}\int |\nabla^2\phi_g(\br)|^2 + \int \bsmear(\br)\phi_g(\br)\dr + G[\phi_g, \vaux, N_e]\,. 
\end{equation}
Thus, 
\begin{equation}
\begin{split}
    \frac{\delta \TE[\vaux, \bsmear, N_e]}{\delta \bsmear(\br)} &= \int \left(\frac{1}{4\pi}\nabla^2\phi_g(\br') + \bsmear(\br') + \frac{\delta G[\phi_g,\vaux, N_e]}{\delta \phi_g(\br')}\right)\frac{\delta \phi_g(\br')}{\delta \bsmear(\br)}\dr + \phi_g(\br) \\
    & = \phi_g(\br)\,,
\end{split}
\end{equation}
where the second line results from Eq.~\ref{SIeq:dLPhi}.

\subsection{Derivation of covariant transformation of $\rho_g$ and $\phi_g$} \label{SIsec:covariantTransformation}
We now derive the covariant transformation of 
$\rho_g$ and $\phi_g$ when the input fields to  $\TE[\vaux, \bsmear, N_e]$ transform as $\vaux^{\prime}=T \vaux$ and $\bsmear^{\prime}=T \bsmear$, where the linear operator $T$ is a representation of the Euclidean symmetry group. Formally, the action of $T$ of a function $f(\br)$ is defined as: $f'(\br') = \int T(\br',\br)f(\br)\dr$. In other words, $T(\br',\br) = \frac{\delta f'(\br')}{\delta f(\br)}$. Using the above definition, we derive the covariance of $\rho_g$ with respect to $(\vaux,\bsmear)$. The derivation for the $\phi_g$ covariance with $(\vaux,\bsmear)$ follows along similar lines. Under Euclidean symmetry group transformations, $\TE$ remains invariant:
\begin{equation}
    \TE[\vaux^{\prime}, \bsmear^{\prime}, N_e]=\TE[\vaux, \bsmear, N_e]\,.
\end{equation}
Now, consider the change (differential) in $\TE$ due to a small variation $\delta \vaux(\br)$ in $\vaux(\br)$. The differentials on both sides of the above equation turn out to be  
\begin{align}
\int \int \frac{\delta \TE[\vaux^{\prime}, \bsmear^{\prime}, N_e]}{\delta \vaux^{\prime}(\br')}\frac{\delta \vaux^{\prime}(\br')}{\delta \vaux(\br)} \delta \vaux(\br) \dr\dr' = & \int \frac{\delta \TE[\vaux, \bsmear, N_e]}{\delta \vaux(\br)}  \delta \vaux(\br) \,\text{d}\,\br\,.
\end{align}
Note that for the left side in the above equation, the term $\int \frac{\delta \vaux'(\br')}{\delta \vaux(\br)}\delta \vaux(\br)\dr$ is nothing but $T\delta \vaux$. Further, we know that $\delta \TE[\vaux', \bsmear', N_e]/\delta \vaux'(\br') = \rho_g[\vaux',\bsmear'](\br')$ and  $\delta \TE[\vaux, \bsmear, N_e]/\delta \vaux(\br) = \rho_g[\vaux,\bsmear](\br)$. Thus, using the bra-ket notation, we can rewrite the above equation as 
\begin{align}\label{eq:covDifferential}
    &\braket{T\delta \vaux}{ \rho_g[\vaux^{\prime},\bsmear^{\prime}]}=\braket{\delta \vaux}{ \rho_g[\vaux,\bsmear]}\notag\,,~\text{or}\\
    &\braket{\delta \vaux}{ T^{\dagger}\rho_g[\vaux^{\prime},\bsmear^{\prime}]}=\braket{\delta \vaux}{ \rho_g[\vaux,\bsmear]}\,,
\end{align}
where in the last step we have used the definition of the adjoint of a linear operator $A$ ($\braket{Af}{g}=\braket{f}{A^{\dagger}g}$). Finally, as the differential relation in Eq.~\ref{eq:covDifferential} is satisfied for any  $\vaux$ perturbation, we get 
\begin{align}\label{eq:covariantTransform}
    &T^{\dagger}\rho_g[\vaux^{\prime},\bsmear^{\prime}]  = \rho_g[\vaux,\bsmear]\notag\,,~\text{or}\\
    &T^{\dagger}\rho_g[T\vaux,T\bsmear] = \rho_g[\vaux,\bsmear]\,.
\end{align}
This proves the covariance of $\rho_g$ with respect to input fields.

\section{Probe functions}
\begin{figure} [htbp!]
    \centering
    \includegraphics[scale=0.7]{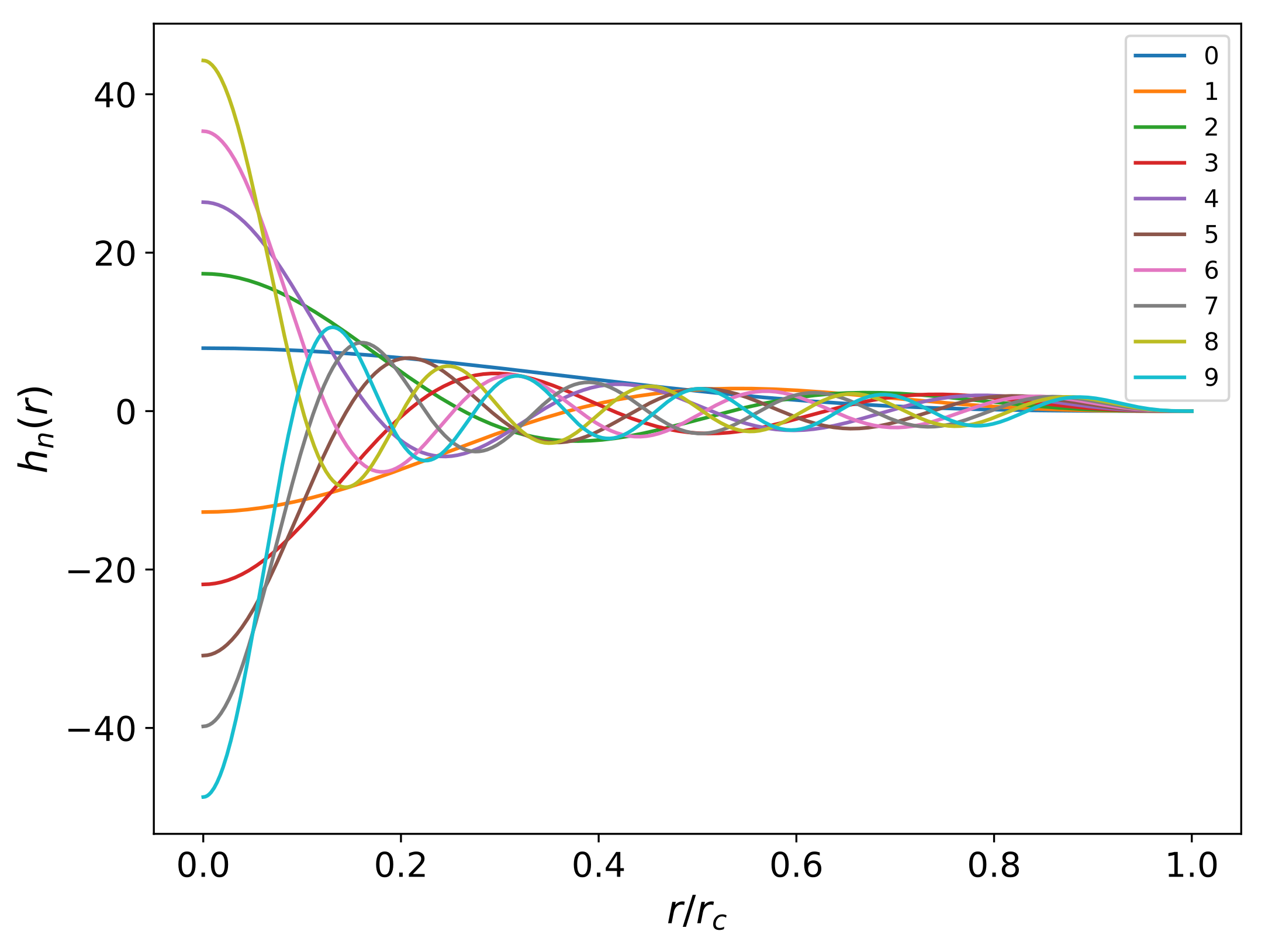}
    \caption{First ten short-ranged orthogonal Bessel (SR-OB) functions.}
    \label{SIfig:bessel}
\end{figure}

Our goal is to use chemical species agnostic probe functions. To that end, we use a combination of short-ranged orthogonal Bessel (SR-OB) functions and long-ranged damped-multipole (LR-DM)
functions for the radial probe functions $f_n(r)$ and $g_n(r)$ (see Eq. 12 in main text). The SR-OB are an extension of the standard spherical Bessel function of first kind~\cite{Kocer2019}. For a given radial cutoff ($r_c$), the $n$-th SR-OB function ($h_n(r)$) is defined as
\begin{equation}
    h_n(r) = \frac{1}{\sqrt{d_n}}\left({j}_n(r) + \frac{e_n}{d_{n-1}}j_{n-1}(r)\right);\quad ~ h_0(r) = j_0(r)\,,
\end{equation}
where 
\begin{equation}
    e_n = \frac{n^2(n+2)^2}{4(n+1)^4+1};\quad d_n = 1- \frac{e_n}{d_{n-1}}; \quad d_0=1\,,~\text{and}
\end{equation}
\begin{equation}
    j_n(r) = (-1)^n \frac{\sqrt{2}\pi}{r_c^{3/2}}\frac{(n+1)(n+2)}{\sqrt{(n+1)^2 + (n+2)^2}} \left[\text{sinc}\left(\frac{r(n+1)\pi}{r_c}\right)+\text{sinc}\left(\frac{r(n+2)\pi}{r_c}\right)\right]\,.
\end{equation}
The $h_n$'s have the following properties: (i) $h_n(r)=0, \,\, \forall r \geq r_c$ (short-ranged); and (ii) $\int h_n(r)h_{n'}(r)r^2\dr = \delta_{nn'}$ (orthogonal). Fig.~\ref{SIfig:bessel} shows the first ten SR-OB functions. We use 20 and 15 SR-OB functions to probe $\vaux$ and $\bsmear$, respectively. We use $r_c$ values of 4 and 6.5 for $\vaux$ and $\bsmear$, respectively.

\noindent The LR-DM functions, $v_{a,b,c,k,r_c,t}(r)$, are given as
\begin{equation}\label{SIeq:LR-DM1}
    v_{a,b,c,k,r_c,t}(r) = \xi_{a,b,c,k}(r)  \, \, s_{r_c,t}(r)\,,
\end{equation}
where 
\begin{equation} \label{SIeq:LR-DM2}
    \xi_{a,b,c,k}(r) = \left(1-\frac{1}{1+ar^b}\right)\frac{\text{exp}(-cr)}{r^k}\,,
\end{equation}
with $a,b,c, k \geq 0$. $s_{r_c,t}(r)$, with $t>0$, is a smooth cutoff function with the following properties: 
\begin{equation}
    \begin{cases} 
      s_{r_c,t}(r) =  1, & r <  r_c\,\frac{t}{t+1} \\
      0 < s_{r_c,t}(r) < 1, & r_c\,\frac{t}{t+1} \leq r < r_c \\
      s_{r_c,t}(r) = 0, & r \geq  r_c\,.
   \end{cases}
\end{equation}
In our work, we use $r_c=24$ and $t=2$ for $s_{r_c,t}$. For the LR-DM probe functions for $\vaux(\br)$, we set $k=0$ (in Eq.~\ref{SIeq:LR-DM2}). This is motivated by the fact that the probe functions for $\vaux$ serves the purpose of being a basis for $\Delta\rho_g(\br)$ and hence should have an exponential decay. Similarly, the probe functions for $\bsmear(\br)$ should serve as a good basis for $\Delta \phi_g(\br)$, which for a charge neutral system has screened multipole decay of the form $\text{exp}(-cr)/r^k$. Accordingly, for the LR-DM probe functions  for $\bsmear(\br)$, we use $k=\{1, 3\}$.

\begin{figure}[t!] \label{SIfig:lrdm}
    \centering
    \begin{subfigure}[b]{0.48\textwidth}
        \centering
        \includegraphics[scale=0.37]{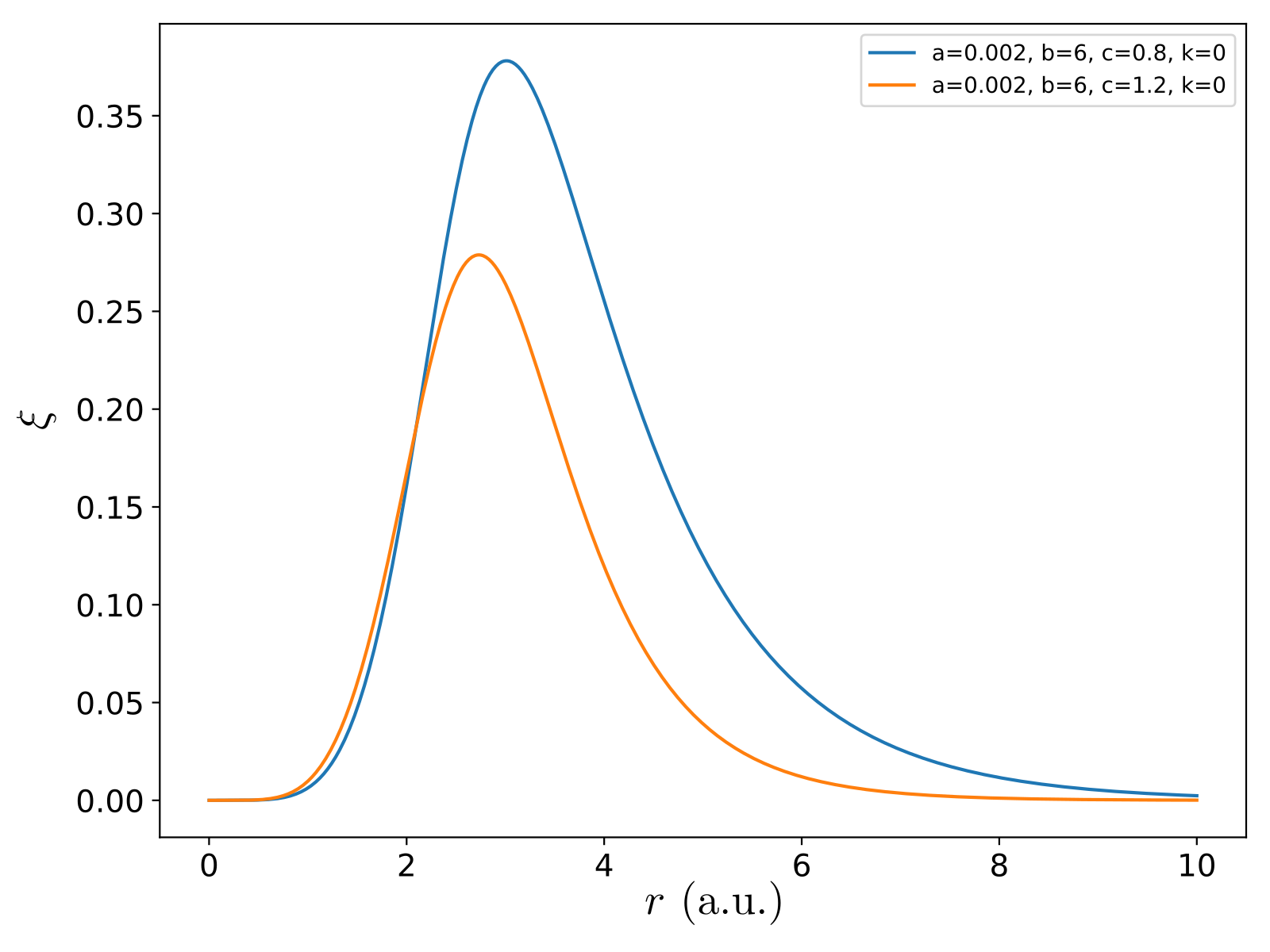}
    \end{subfigure}%
    \begin{subfigure}[b]{0.48\textwidth}
        \centering
        \includegraphics[scale=0.37]{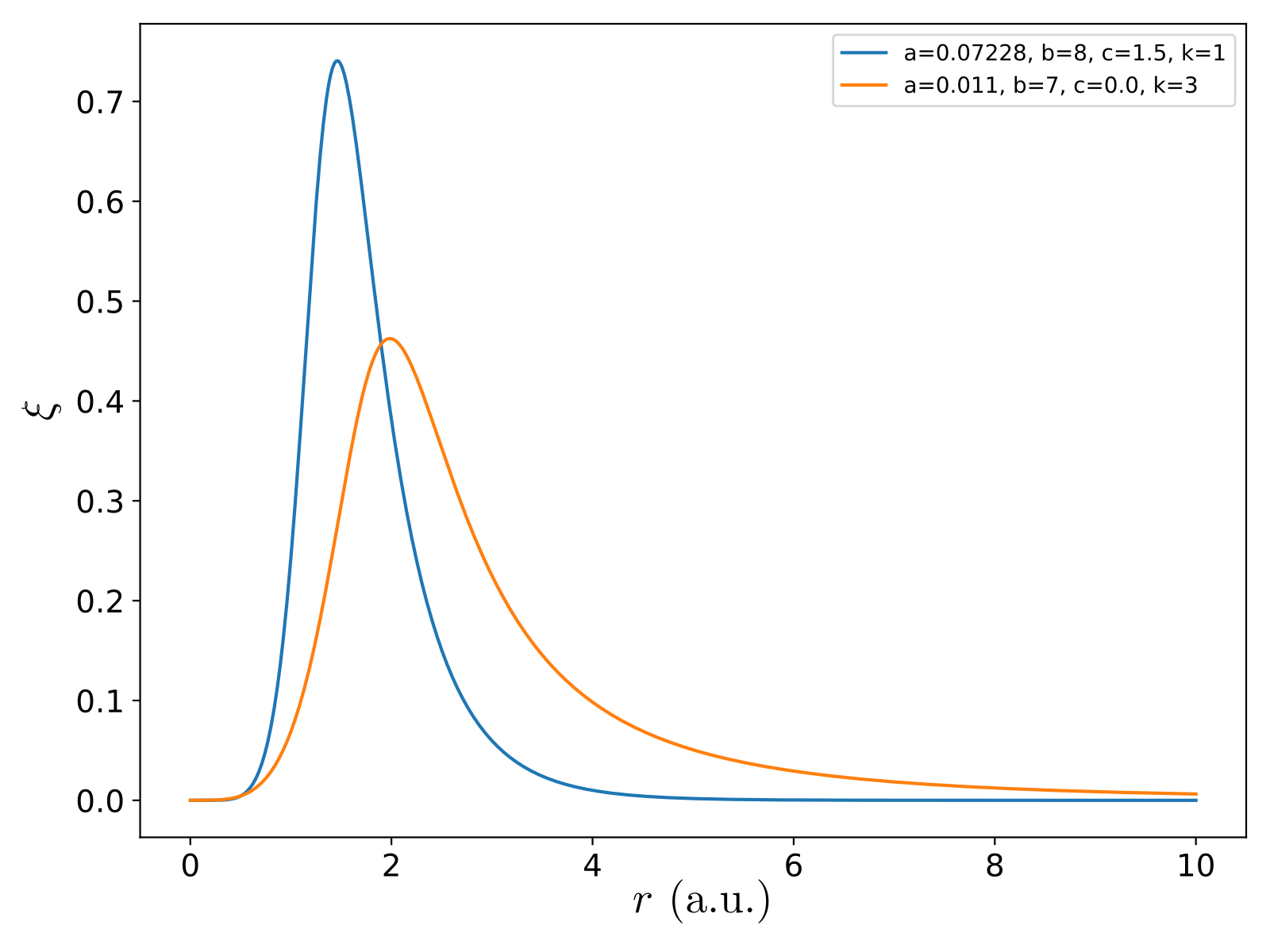}
    \end{subfigure}
    \caption{Long-ranged damped multipole (LR-DM) functions used to probe $\vaux$ (left) and $\bsmear$ (right).}
\end{figure}

\section{Form of nuclear charge distribution}
We recap that our auxiliary nuclear charge takes the form $\bsmear(\br) = \sum_I b_I(\br)$, where $ b_I(\br) = -Z_I \bunit(|\br-\bR_I|; r_c)$. Here, $\bunit(r; r_c)$ is a compactly supported positive unit charge distribution with a cutoff radius of $r_c$. The form of $\bunit(r;r_c)$ we employ is borrowed from Ref.~\cite{Pask2012} and is given as
\begin{equation} \label{SIeq:bu}
    \bunit(r;r_c) = \begin{cases}
    \frac{-21 (r-r_c)^3 (6r^2 + 3r r_c + r_c^2)}{5 \pi r_c^8},  & 0\leq r \leq r_c, \\
    0, & r > r_c\,.
    \end {cases}
\end{equation}
The potential corresponding to $\bunit(r,r_c)$, is given as
\begin{equation} \label{SIeq:vu}
    v_{\text{u}}(r;r_c) = \begin{cases}
    \frac{9r^7 - 30r^6r_c + 28r^5r_c^2 - 14r^2r_c^5 +12r_c^7}{5r_c^8}, &0\leq r \leq r_c \\
    \frac{1}{r}, &r > r_c\,.
    \end{cases}
\end{equation}
Using the above, we can simplify $\vsmear(\br) = \int \frac{\bsmear(\br')}{|\br-\br'|}dr'$ as
\begin{equation}
    \vsmear(\br) = -\sum_I Z_I v_{\text{u}}(|\br-\bR_I|;r_c)\,.
\end{equation}
An $r_c < \frac{1}{2} \min\left(|\bR_I-\bR_J|\right)$ (i.e., less than half the minimum pairwise distance between atoms) ensures non-overlapping nuclear charges. In this work, we use $r_c=0.5$ a.u.  

\section{Reference field data} \label{SIsec:refField}
In this section, we discuss how the reference field, $\Delta \rho_g(\br)$ and $\Delta \phi_g(\br)$, used in our training and testing are represented. For numerical efficiency and compact representation, we represent the fields in the corresponding probe function basis. Let's denote $\eta(\br)$ to be a generic representation for $\Delta \rho_g(\br)$ and $\Delta \phi_g(\br)$. Similarly, let's denote $p_i(\br)$ as a generic representation of the probe function basis (i.e., for $u_i(\br)$ and $w_i(\br)$ defined in Eq. 13 of main text). We define an approximation of $\eta(\br)$ in the $p_i(\br)$ basis, denoted by $\widetilde{\eta}(\br)$, as
\begin{equation}
    \widetilde{\eta}(\br) = \sum_I \sum_i c_{I,i} p_i(\br-\br_I)\,.
\end{equation}
The linear coefficients $c_{I,i}$ can be obtained  through a minimization of the difference between $\eta(\br)$ and $\widetilde{\eta}(\br)$ in an appropriate norm. We evaluate them through the following minimization
\begin{equation}
    \min_{\{c_{I,i}\}} \int \int \left(\eta(\br)-\widetilde{\eta}(\br)\right)\left(a \delta(\br-\br') + \frac{b}{|\br-\br'|} \right)  \left(\eta(\br')-\widetilde{\eta}(\br')\right)\dr\dr'\,.
\end{equation}
The solution to the above takes the form of the following linear system of equations
\begin{equation}
    (a \bS + b\bJ)\bc = a \bt_S + b\bt_J\,,
\end{equation}
where $\bS$ and $\bJ$ are the overlap and Coulomb matrix for the $p_i$ basis, given as
\begin{equation} \label{SIeq:SJMatrix}
    S_{I,i}^{J,j} = \int p_i(\br-\bR_I)p_j(\br-\bR_J)\,,\quad J_{I,i}^{J,j}=\int \int \frac{p_i\left(\br-\bR_I\right)p_j\left(\br'-\bR_J\right)}{|\br-\br'|}\dr\dr'\,.
\end{equation}
The vectors $\bt_S$ and $\bt_J$ are defined as
\begin{equation}
    b_S^{I,i} = \int \eta(\br) p_i(\br-\bR_I)\dr\,, \quad b_{J}^{I,i} = \int \int \frac{\eta(\br)p_i\left(\br'-\bR_I\right)}{|\br-\br'|}\dr\dr'\,. 
\end{equation}
Using $b=0$ results in the familiar $L_2$ (or least square) projection. Using $a=0$ leads to minimization in the Coulomb norm. In this work, we use $a=1,b=1$ for the projection of $\Delta \rho_g(\br)$ and $a=1, b=0$ for the projection of $\Delta \phi_g(\br)$. 

\section{Evaluation of field loss} \label{SIsec:fieldloss}
Recall that our loss function takes the form (see Eq. 18 in main text),
\begin{equation} \label{SIeq:loss}
\begin{split}
    \mathcal{L} &= \frac{t_E}{M} \sum_{\alpha=1}^M \left(\DTE^{(\alpha)}_{\text{ref}}-\DTE^{(\alpha)}\right)^2 + \frac{t_F}{M}  \sum_{\alpha=1}^{M} \left[\sum_I \frac{1}{N_a}\left(\norm{\bff^{(\alpha)}_{I,\text{ref}}-\bff^{(\alpha)}_I}^2 \right) \right] \\ 
    & + \frac{t_{\rho}}{M} \sum_{\alpha=1}^{M}\norm{\Delta\rho_{g,\text{ref}}^{(\alpha)}(\br) - \Delta\rho_g^{(\alpha)}(\br)}_{\text{J}}^2
    + \frac{t_{\phi}}{M} \sum_{\alpha=1}^{M}\norm{\Delta\phi_{g,\text{ref}}^{(\alpha)}(\br) - \Delta\phi_g^{(\alpha)}(\br)}^2_{\text{S}}\,,
\end{split}
\end{equation}
where $\alpha$ indexes the $M$ training samples; $t_E$, $t_F$, $t_\rho$, and $t_\phi$ are the weights assigned to the energy, force, density, and potential loss terms; $\norm{\cdot}$ is the  Eucledian norm, $\norm{f(\br)}_{\text{S}} = \sqrt{\int (f(\br))^2\dr}$ is the $L_2$ norm for a continuous function; and $\norm{f(\br)}_{\text{J}} = \sqrt{\int \int \frac{f(\br)f(\br')}{|\br-\br'|}\dr\dr'}$ is the Coulomb norm. In the above, the third and fourth term are the field loss terms corresponding to the density and potential, respectively. Using a 3D quadrature grid to evaluate the field loss terms can be computationally expensive, especially when it is done in every epoch in the training. This cost can be alleviated by using the fact that both the reference and the predicted fields are represented in the same basis. To elaborate, consider the $\bc_{\alpha}$ and $\widetilde{\bc}_{\alpha}$ be the linear coefficients for $\Delta \rho_{g,\text{ref}}^{(\alpha)}(\br)$ and $\Delta \rho_{g}^{(\alpha)}(\br)$, respectively, in the $u_i(\br)$ basis. Thus, we can rewrite the density loss $\norm{\Delta\rho_{g,\text{ref}}^{(\alpha)}(\br) - \Delta\rho_g^{(\alpha)}(\br)}_{\text{J}}^2 = \bc_{\alpha}^{T} \bJ \widetilde{\bc}_{\alpha}$. Similarly, denoting $\bd_{\alpha}$ and $\widetilde{\bd}_{\alpha}$ as the linear coefficients for $\Delta \phi_{g,\text{ref}}^{(\alpha)}(\br)$ and $\Delta \phi_{g}^{(\alpha)}(\br)$, respectively, in the $w_i(\br)$ basis, the potential loss can be written as $\norm{\Delta\phi_{g,\text{ref}}^{(\alpha)}(\br) - \Delta\phi_g^{(\alpha)}(\br)}^2_{\text{S}}=\bd_{\alpha}^T\bS \widetilde{\bd}_{\alpha}$. In other words, the field loss terms can be efficiently recast into matrix-vector products.